\documentclass[aps,prc,reprint,showpacs,groupedaddress,onecolumn]{revtex4-1}

\usepackage{graphicx,color} % Include figure files 
\usepackage{dcolumn}  % Align table columns on decimal point  
\usepackage{bm}       % bold math 
\usepackage{amsmath}

\begin{document}
\newcommand{\vv}[1]{{$\bf {#1}$}}
\newcommand{\ul}[1]{\underline{#1}}
\newcommand{\vvm}[1]{{\bf {#1}}}
\def\bsigma{\mbox{\boldmath$\sigma$}}
\def\btau{\mbox{\boldmath$\tau$}}
\newcommand{\IR}[1]{\textcolor{red}{#1}}

\title{Separable Representation of Phenomenological Optical Potentials \\ of Woods-Saxon Type} 

\author{L.~Hlophe$^{(a)}$}
\email{lh421709@ohio.edu}
\author{Ch.~Elster$^{(a)}$}
\email{elster@ohio.edu}
\author{ R.C.~Johnson$^{(b)}$,
N.J~Upadhyay$^{(c)}$, F.M.~Nunes$^{(c)}$, G.~Arbanas$^{(d)}$, V.~Eremenko$^{(a,f)}$
J.E.~Escher$^{(e)}$, I.J.~Thompson$^{(e)}$
}

\affiliation{(a)Institute of Nuclear and Particle Physics,  and
Department of Physics and Astronomy,  Ohio University, Athens, OH 45701 \\
(b) Department of Physics, University of Surrey, Guildford GU2 7XH, UK  \\
(c) National Superconducting Cyclotron Laboratory and Department of Physics and Astronomy, Michigan State University, East Lansing, MI 48824, USA \\
(d) Nuclear Science and Technology Division, Oak Ridge National Laboratory, Oak Ridge, TN 37831, USA \\
(e) Lawrence Livermore National Laboratory L-414, Livermore, CA 94551, USA
(f) D.V. Skobeltsyn Institute of Nuclear Physics, M.V. Lomonosov
Moscow State University, Moscow, 119991, Russia
}

\collaboration{The TORUS Collaboration }
\noaffiliation

\date{\today}

\begin{abstract}
{\bf Background:} {One important ingredient for many applications of nuclear physics to astrophysics,
nuclear energy,  and stockpile stewardship are cross sections for reactions of neutrons with rare isotopes.
Since direct measurements are often not feasible, indirect methods, e.g. (d,p) reactions, should be used.}
Those (d,p) reactions may be viewed as three-body reactions and described with Faddeev techniques.
{\bf Purpose:} Faddeev equations in momentum space have a long tradition of
utilizing separable interactions in order to arrive at sets of coupled integral
equations in one variable. While there exist several separable representations
for the nucleon-nucleon interaction, the optical potential between a neutron
(proton) and a nucleus is not readily available in separable form. The purpose
of this paper is to introduce a separable representation for complex phenomenological
optical potentials of Woods-Saxon type.
{\bf Results:} Starting from a global optical potential, a separable
representation thereof is introduced based on the Ernst-Shakin-Thaler (EST) scheme. This
 scheme is generalized to non-hermitian potentials. 
Applications to n$+^{48}$Ca, n$+^{132}$Sn and n$+^{208}$Pb are investigated for 
energies from 0 to 50~MeV and the quality of the representation is examined.
{\bf Conclusions}: We find a good description of the on-shell t-matrix for all systems with rank up to 5. The required rank depends inversely on the angular momentum. The resulting separable interaction exhibits a different off-shell behavior compared to the original potential, reducing the high momentum contributions.
\end{abstract}


\maketitle

\section{Introduction}
\label{intro}

A variety of applications of nuclear physics require the understanding of neutron
capture on unstable nuclei. Due to the short lifetimes involved, direct measurement
are currently not possible, and thus indirect methods using (d,p) reactions have been
proposed (see e.g Refs.~\cite{RevModPhys.84.353,jolie,Kozub:2012ka}). 
Single neutron transfer (d,p) reactions
have also been the preferred tool to study shell evolution in nuclear structure (see e.g.
Refs.~\cite{Schmitt:2012bt,Jones:2011kp}). 
In all these cases, a reliable reaction theory is a critical ingredient.

Scattering and reaction processes involving deuterons either as projectile or
as target are perhaps the most natural three-body problem in the realm of nuclear reactions. The binding energy of the deuteron is so small that its root-mean-square radius is significantly larger than the range of the force. That means, when a deuteron interacts with a compact nucleus, one may expect that it will behave like a three-body system consisting of proton $p$, a neutron $n$, and a nucleus $A$. The obvious three-body reactions are elastic scattering, rearrangement and breakup processes. 
In order to describe those processes on the same footing, deuteron-nucleus
scattering should be treated as a three-body problem with a three-body Hamiltonian
governing the dynamics.  This three-body Hamiltonian contains the well understood nucleon-nucleon
(NN) interaction as well as the effective interactions between the nucleons and the target
($nA$ and $pA$). It is common for these nucleons-nucleus interactions to take  phenomenological optical potentials which fit a large body of elastic scattering data~\cite{Varner:1991zz,Weppner:2009qy,Koning:2003zz,Becchetti:1969zz}.

The application of momentum space Faddeev techniques to nuclear reactions has been pioneered in Ref.~\cite{Deltuva:2009fp}, and successfully applied to (d,p) reactions for light nuclei~\cite{Deltuva:2009cr}. However, when extending these
calculations to heavier nuclei~\cite{Nunes:2011cv,Upadhyay:2011ta}, it becomes
 apparent that
techniques employed for incorporating the Coulomb interaction in Faddeev-type
calculations of reactions with light nuclei can not readily
be extended to the heaviest nuclei. Therefore, a new method for treating 
(d,p) reactions with the exact inclusion of the Coulomb force
 as well as target excitation was formulated in Ref.~\cite{Mukhamedzhanov:2012qv}. 
This new approach relies on a separable representation of the interparticle forces. 

Separable representations of the forces between constituents forming the subsystems in a Faddeev approach have a long tradition in few-body physics. There is a large body of work on separable representations of 
 NN interactions (see e.g. Refs.~\cite{Haidenbauer:1982if,Haidenbauer:1986zza,Berthold:1990zz,Schnizer:1990gf,Entem:2001it}) or  meson-nucleon interactions~\cite{Ueda:1994ur,Gal:2011yp}.
In the context of describing light nuclei like $^6$He~\cite{Ghovanlou:1974zza}
and $^6$Li~\cite{Eskandarian:1992zz} in a three-body approach, separable interactions have been successfully used. 
A separable nucleon-$^{12}$C optical potential was proposed in
Ref.~\cite{MiyagawaK}, consisting of a rank-1 Yamaguchi-type form factor fitted to
the positive energies and a similar term describing the bound states in the
nucleon-$^{12}$C configuration. However,
we are not aware of any systematic work along this line for heavy nuclei, 
for which  excellent phenomenological descriptions exist in terms of
Woods-Saxon functions~\cite{Varner:1991zz,Weppner:2009qy,Koning:2003zz,Becchetti:1969zz}. 
For applications to (d,p) reactions  there is a need for a procedure for deriving a separable
representation that 
is sufficiently general for a variety of
nucleon-nucleus optical potentials, as well as over a wide range of nuclei and energies, 
so that one can take advantage of the already existing extensive work on 
phenomenological optical potentials.

The separable representation of two-body interactions suggested by
Ernst-Shakin-Thaler~\cite{Ernst:1973zzb} (EST) seems well suited for achieving
this goal. We note that this EST approach
has been successfully employed to represent NN potentials~\cite{Haidenbauer:1982if,Haidenbauer:1986zza}. 
However, the EST representation as derived in Ref.~\cite{Ernst:1973zzb},
 though allowing energy dependence of the potentials~\cite{Ernst:1974zzb,Pearce:1987zz},
assumes that they are Hermitian. 
Therefore, the EST approach needs to be generalized in order to be applicable for
optical potentials which are complex.

In Section~\ref{formal} we present the  generalization of the EST approach to
non-Hermitian potentials which is needed so that the potential as well as the
transition matrix fulfills the reciprocity theorem. First we explicitly show for a 
rank-1 separable potential the required redefinition of the separable ansatz of
Ref.~\cite{Ernst:1973zzb}, and then generalize to separable potentials of arbitrary
rank. Since all our calculations are carried out in momentum space, we sketch the
explicit procedure to obtain separable transition matrices.  
This procedure follows closely the one laid out in Ref.~\cite{Ernst:1973zzb},
and we will refer to it as EST scheme. 

In  Section~\ref{ESTch89} we  present the results for the separable  representations 
of optical potentials for $^{48}$Ca, $^{132}$Sn, and $^{208}$Pb based on the CH89~\cite{Varner:1991zz} 
phenomenological optical potential. Note that the definition of the optical potential in coordinate space, as well as details on
the Fourier transform and partial wave decomposition are presented in the Appendix. In 
Section~\ref{offshell},  we  investigate the off-shell behavior of the 
separable representations and compare it to the original potential. 
Finally, we  summarize our findings in Section~\ref{summary}.

%%%%%%%%%%%%%%%%%%%%%%%%%%%%%%%%%%%%%%%%%%%%%%%%%%%%%%%%%%%%%%%%%

\section{Formal Considerations}
\label{formal}

A scheme for constructing separable potentials from the solution of
a Lippmann-Schwinger (LS) equation with arbitrary hermitian potentials
was suggested by Ernst, Shakin, and Thaler~\cite{Ernst:1973zzb}. This
separable potential is designed to represent the scattering matrix (or
equivalently the scattering phase shifts) over a chosen energy range 
with the same quality as the original potential. Furthermore, this EST
scheme provides a well-defined prescription for increasing the accuracy
of the representation by increasing the rank of the separable potential.

The basic idea of the Ernst-Shakin-Thaler (EST) scheme 
for constructing a separable
representation of the two-body transition amplitude is that one selects a fixed
number of energy points in the energy interval in which the separable
potential shall represent the scattering matrix. At these chosen points,
the on-shell as well as half-shell  t-matrices of the original potential are
identical to those obtained with the separable potential. The corresponding
half-shell t-matrices at these points 
then serve as form-factors of the separable representation. The number
of these fixed points gives the rank of the separable potential.

The  EST~\cite{Ernst:1973zzb,Ernst:1974zzb} scheme has been 
successfully applied to construct separable representations of 
several NN potentials~\cite{Haidenbauer:1982if,Haidenbauer:1986zza} defined below the
pion production threshold, i.e. real potentials. Potentials intended to
describe the scattering of neutrons or protons from nuclei are in general
complex as a result of reactions channels not explicitly taken into account.
In the following we show that the formulation given in
Ref.~\cite{Ernst:1973zzb} is not suitable for complex potentials, since it is not
compatible with the reciprocity theorem.  We illustrate how the requirement of
time-reversal invariance leads to a modification of the definition of the separable
potential.

\subsection{Separable Complex Potentials of Rank-1 and Time Reversal}
\label{subEST}

For applications to the theory of nuclear reactions it is convenient to arrange that all potential operators $U$ satisfy
\begin{equation}
{\cal K} U {\cal K}^{-1} = U^\dagger,
\label{eq:2.2.1}
\end{equation} 
where $\cal{K}$ is the time reversal operator appropriate to the system. This condition guarantees that the $S$-matrix corresponding to $U$ is symmetric  and that reaction amplitudes constructed from these potentials satisfy convenient reciprocity relations.

When  $U$ is a central potential in the space of a spinless particle,  $\cal{K}$ can be chosen 
to be the antilinear complex conjugation operator ${\cal K}_0$, which in the 
coordinate space basis $|{\bf r}\rangle$ is
 defined by 
\begin{equation}
{\cal K}_0\; \alpha \,|{\bf r}\rangle = \alpha^* ( {\cal K}_0 |{\bf r}\rangle) =
\alpha^*|{\bf r}\rangle,
\label{eq:2.2.2}
\end{equation}
and from which we deduce ${\cal K}_0 |{\bf p}\rangle = |-{\bf p}\rangle.$
 Note that for this particular  ${\cal K}$ we have $({\cal K}_0)^{-1}={\cal K}_0$.
 
We first consider a Hermitian interaction $v$ acting in a partial wave with angular momentum $l$. 
It was shown in Ref.~\cite{Ernst:1973zzb} that  a rank-1 separable 
potential leading to a scattering wave function that is identical to that of potential $V$ (defined via
a Hamiltonian $H=H_0 +V$) at
a specific energy $E_{k_E}$  (support point) is given as
\begin{equation}
{\bf V}(E_{k_E}) = \frac{ v | f_{l,k_E}\rangle \langle  f_{l,k_E}|v}
           {\langle f_{l,k_E} |v| f_{l,k_E}\rangle} \equiv v |
f_{l,k_E}\rangle {\hat \lambda} \langle f_{l,k_E}|v \; .
\label{eq:2.2.3}
\end{equation}
Here $| f_{l,k_E}\rangle$ is the regular radial scattering wave function, unique within an 
overall constant, for $v$ at energy $E_{k_E}$, 
$v | f_{l,k_E}\rangle$ is the form factor, and $({\hat \lambda})^{-1} = \langle f_{l,k_E}
|v| f_{l,k_E}\rangle $ the strength parameter. 

For a Hermitian $v$ the radial function $f_{l,k_E}$ can be taken to be real in both coordinate 
and momentum space and the strength parameter ${\hat \lambda}$ is real. As a result the 
non-local potential  ${\bf V}$ of Eq.~(\ref{eq:2.2.3}) is a Hermitian symmetric matrix in 
both momentum and coordinate space and satisfies ${\cal K}_0 V {\cal K}_0=V ^{\dagger}$.

If however $v$ is a complex potential, or more generally a non-hermitian operator $u$ the 
radial functions $f_{l,k_E}$ are no longer real and if $v$ is simply replaced by $u$ in 
Eq.~(\ref{eq:2.2.3}) the resulting rank-1 non-local potential $U$ will neither be
 Hermitian nor satisfy ${\cal K}_0 U {\cal K}_0=U ^{\dagger}$.

To remedy this situation for a non-Hermitian potential $u$ we replace the definition 
of Eq.~(\ref{eq:2.2.3}) by
\begin{equation}
{\bf U}(E_{k_E}) \equiv \frac{ u | f_{l,k_E}\rangle \langle  f^*_{l,k_E}|u}
           {\langle f^*_{l,k_E} |u| f_{l,k_E}\rangle} \equiv u |
f_{l,k_E}\rangle {\hat \lambda} \langle f^*_{l,k_E}|u \; ,
\label{eq:2.2.4}
\end{equation}
 where now the strength parameter is defined by $({\hat \lambda})^{-1} = \langle  f^*_{l,k_E}
|u| f_{l,k_E}\rangle $.

Here $f_{l,k_E}(r)$ is the unique regular radial wave function corresponding to $u$ and $f^*_{l,k_E}(r)$ is the unique regular radial wavefunction corresponding to $u^*$. By a suitable choice of arbitrary normalization constants we can arrange that $f^*_{l,k_E}(r)$ is simply the complex conjugate of $f_{l,k_E}$ and hence ${\cal K}_0|f_{l,k_E}\rangle=|f^*_{l,k_E}\rangle $.
   
 If $u$ satisfies ${\cal K}_0 u {\cal K}_0=u^{\dagger}$ the definition of Eq.~(\ref{eq:2.2.4})
 gives  a symmetric complex potential matrix that satisfies 
 \begin{eqnarray}
{\cal K}_0{\bf U}(E_{k_E}){\cal K}_0 &=& ({\cal K}_0 u |
f_{l,k_E}\rangle) ({\hat \lambda})^* (\langle f^*_{l,k_E}|u {\cal K}_0)\nonumber \\
&=&u^{\dagger} |
f^*_{l,k_E}\rangle ({\hat \lambda})^* \langle f_{l,k_E}|u^{\dagger}\nonumber \\
&=&U^{\dagger},
\label{eq:2.2.5}
\end{eqnarray}
where the round brackets mean that ${\cal K}_0$ here acts only on the quantities within the brackets.

For a general energy $E$ and arbitrary potential ${\cal V}$ we define an operator $t(E)$ 
as the solution of
\begin{eqnarray}
t(E)={\cal V}+{\cal V}g_0 (E) t(E).
\label{eq:2.2.6}
\end{eqnarray}
For the potential given by Eq.~(\ref{eq:2.2.4}) 
we then obtain  a rank-1 separable t-matrix $t(E)$ in a given partial wave with matrix elements
\begin{eqnarray}
\langle p'|t(E)| p\rangle &=& \frac{\langle p'|u | f_{l,k_E}\rangle \langle
f^*_{l,k_E}|u|p\rangle }{ \langle f^*_{l,k_E} |u -ug_0(E)u|
f_{l,k_E}\rangle } .
\label{eq:2.2.6a}
\end{eqnarray}
Introducing $ t (p',k_E, E_{k_E})=\langle f^*_{l,k_E}|u| p' \rangle$,  
and  $ t (p,k_E, E_{k_E})=\langle p |u| f_{l,k_E} \rangle$, the partial wave
t-matrix element  $\langle p'|t(E)| p\rangle$ can be written as
\begin{eqnarray}
\langle p'|t(E)| p\rangle &=& \frac {t (p',k_E, E_{k_E}) \;  t(p,k_E, E_{k_E})}{\langle f^*_{l,k_E}|
u(1-g_0(E) u) |f_{l,k_E}\rangle} \equiv t (p',k_E, E)\;  \tau(E)\;  t(p,k_E, E), 
\label{eq:2.2.7}
\end{eqnarray}
where
\begin{equation}
\tau^{-1}(E) = \langle f^*_{l,k_E}|u(1-g_0(E) u) |f_{l,k_E}\rangle .
\label{eq:2.2.8}
\end{equation}
The scattering wavefunction $|f_{l,k_E}\rangle$  satisfies
$|f_{l,k_E}\rangle = |k_E\rangle + g_0(E_{k_E}) u |f_{l,k_E}\rangle$. Using this
we find $(\tau(E_{k_E}))^{-1}=\langle k_E|u|f_{l,k_E}\rangle=t (k_E,k_E, E_{k_E})$, 
and hence
\begin{equation}
\langle p'|t(E_{k_E})| p\rangle  = \frac {t (p',k_E, E_{k_E}) \;
t(p,k_E, E_{k_E})} { t (k_E ,k_E, E_{k_E})},
\label{eq:2.2.9}
\end{equation} 
where angular momentum indices are omitted.

On the energy shell, i.e. for $p \rightarrow k_E$ and $p' \rightarrow k_E$, 
the separable
t-matrix of Eq.~(\ref{eq:2.2.9}) agrees with the t-matrix evaluated with the original
potential $u$, as it should.
For any general $E = k_0^2/2\mu$ the function $\tau(E)$ of Eq.~(\ref{eq:2.2.7}) is explicitly
calculated as
\begin{eqnarray}
\lefteqn{\tau(E)^{-1} = t(k_E,k_E,E_{k_E})}\cr
&&+ 2\mu \left[ {\cal P}\int dp p^2 \frac{ t(p,k_E,E_{k_E})t(p,k_E,E_{k_E})}{k_E^2 -p^2} - 
           {\cal P} \int dp p^2 \frac{t(p,k_E,E_{k_E})t(p,k_E,E_{k_E})}{k_0^2 -p^2} \right] \cr
&& + i \pi \mu \Big[ k_0 t(k_0,k_E,E_{k_E})t(k_0,k_E,E_{k_E}) -k_E \;
t(k_E,k_E,E_{k_E})t(k_E,k_E,E_{k_E}) \Big] \; .
\label{eq:2.2.10}
\end{eqnarray}
Here the half-shell t-matrices $t(p,k_E,E_{k_E})$ are the momentum space
solutions of a standard LS equation at the scattering energy $E_{k_E}$. 
 
Thus, the rank-1 separable potential as given in Eq.~(\ref{eq:2.2.4}) leads to the
desired rank-1 separable t-matrix, which fulfills the reciprocity theorem.

\subsection{Separable Complex Potentials of Arbitrary Rank}
\label{subgeneralEST}

It remains to generalize the  above formulation of a rank-1 separable complex 
potential to one of arbitrary rank. 
In analogy to the procedure followed in Ref.~\cite{Ernst:1973zzb} we 
define a complex separable potential of arbitrary rank in a given partial wave  
as
\begin{equation}
{\bf U}  = \sum_{i,j} u| f_{l,k_{E_i}} \rangle \langle f_{l,k_{E_i}} |M | f^*{_l,k_{E_j}}\rangle
\langle f^*_{l,k_{E_j}}|u .
\label{eq:2.3.1}
\end{equation}
Here $f_{l,k_{E_i}}$ is the unique regular radial wave function corresponding to the complex
potential $u$ and 
asymptotic energy $E_i$, and $f^*_{l,k_{E_i}}$ the unique regular radial wave function
corresponding to $u^*$. Note that $u$ may also be energy dependent.  
The  matrix $M$ is  defined and constrained by
\begin{equation}
\delta_{ik}=\sum_j \langle f_{l,k_{E_i}}|M| f^*_{l,k_{E_j}}\rangle \langle  f^*_{l,k_{E_j}} |
u| f_{l,k_{E_k}}\rangle =
 \sum_j \langle f^*_{l,k_{E_i}} | u | f_{l,k_{E_j}} \rangle \langle f_{l,k_{E_j}}
|M|f^*_{l,k_{E_k}}\rangle.
\label{eq:2.3.2}
\end{equation}
The corresponding separable partial wave t-matrix must be of the form
\begin{equation}
t(E) =  \sum_{i,j} u| f_{l,k_{E_i}}\rangle \tau_{ij}(E) \langle f^*_{l,k_{E_j}}|u \; ,
\label{eq:2.3.3}
\end{equation}
where angular momentum indices are omitted for simplicity of notation.
The coefficient matrix $\tau_{ij}(E)$ is constrained by
\begin{equation}
\sum_i \langle f^*_{l,k_{E_n}}|u - u g_0(E) u| f_{l,k_{E_i}}\rangle \; \tau_{ij}(E) = \delta_{nj}\; ,
\label{eq:2.3.4}
\end{equation}
and
\begin{equation}
\sum_j \tau_{ij}(E) \; \langle f^*_{l,k_{E_j}}|u -u g_0(E) u| f_{l,k_{E_k}}\rangle = 
\delta_{ik} \;.
\label{eq:2.3.5}
\end{equation}
For $i=j=1$  we recover the expressions for the rank-1 potential of the previous section.

For the explicit calculation of the matrix   $\tau_{ij}(E)$, we define a matrix
\begin{equation}
R_{ij}(E) \equiv \langle f^*{_l,k_{E_i}}|u -u g_0(E) u| f_{l,k_{E_j}}\rangle ,
\label{eq:2.3.6}
\end{equation}
so that the  condition of Eq.~(\ref{eq:2.3.5}) reads
\begin{equation}
\sum_j \tau_{ij}(E) R_{jk}(E) = \delta_{ik},
\label{eq:2.3.7}
\end{equation}
from which follows
\begin{equation}
\tau_{ij} (E) = \left( R(E) \right)^{-1}_{ij}.
\label{eq:2.3.8}
\end{equation}

Using again that $ t (p',k_{E_i}, E_{i})=\langle f^*_{l,k_{E_i}}|u| p' \rangle$, and
$ t (p,k_{E_i}, E_{i})=\langle p |u| f_{l,k_{E_i}} \rangle$,
the matrix elements $R_{ij}$ are calculated  in momentum space. 
With $E=k_0^2/2\mu$, the diagonal matrix elements are given as
\begin{eqnarray}
R_{ii}(E)&=& \langle f^*_{l,k_{E_i}}|u-ug_0(E) u| f_{l,k_{E_i}}\rangle \cr
&=& t(k_{E_i},k_{E_i};E_i) +
 2\mu {\cal P} \int dp p^2 \frac{t(p,k_{E_i};E_i) t(p,k_{E_i};E_i)} {k_{E_i}^2 -p^2} \cr
 & & -i\pi k_{E_i} t(k_{E_i},k_{E_i};E_i)t(k_{E_i},k_{E_i};E_i) \cr
&&- 2\mu {\cal P} \int dp p^2 \frac{t(p,k_{E_i};E_i)t(p,k_{E_i};E_i)}{k_0^2 -p^2}  \cr
& & +i\pi k_0 t(k_0,k_{E_i};E_i) t(k_0,k_{E_i};E_i).
\label{eq:2.3.9}
\end{eqnarray}
The half-shell t-matrices $t(p,k_{E_i};E_i)$ are the momentum space solutions of a standard
LS equation at the support points $E_i$. At these energies $E_i$,  the on-shell
t-matrix elements calculated from the separable potential agree with those calculated from the original complex potential $u$. 

The off-diagonal elements of the matrix $R_{ij}(E)$ are given by
\begin{eqnarray}
R_{ij}(E)&=& \langle f^*_{l,k_{E_i}}|u-ug_0(E) u|f_{l,k_{E_j}}\rangle \cr
&=& t(k_{E_j},k_{E_i};E_i) \cr
&& +2\mu{\cal P} \int dp p^2 
 \frac{t(p,k_{E_i};E_i) t(p,k_{E_j};E_j)}{k_{E_j}^2 -p^2} -i\pi \mu k_{E_j}
 t(k_{E_j},k_{E_i};E_i) t(k_{E_j},k_{E_j};E_j) \cr
&&- 2\mu {\cal P} \int dp p^2 \frac{ t(p,k_{E_i};E_i) t(p,k_{E_j};E_j)}
 {k_0^2 -p^2} +i\pi \mu k_0 t(k_0,k_{E_i};E_i)t(k_0,k_{E_j};E_j).
\label{eq:2.3.10}
\end{eqnarray}

The matrix elements $R_{ij}(E)$, where $i,j$ determine the rank of the separable
representation,  are calculated at a fixed number of energy support points $E_i$. The elements
$\tau_{ij}(E)$ are then obtained by solving a system of linear equations 
${\bf R} \btau = {\bf 1}$. For this to be successful, the inverse of $R_{ij}(E)$ must exist.
For calculating the form factors, the half-shell t-matrices at the support points $E_i$  
are calculated by solving a LS equation with the momentum space Woods-Saxon potential derived in
Appendix~\ref{appendixA}.  For calculating the matrix elements $R_{ij}(E)$  we use the cubic Hermite spline
interpolation of Ref.~\cite{Huber:1996td} to obtain the values at the required momenta. Since
spline functions are by design linearly independent, the representation of the half-shell
t-matrices in this basis guarantees that the matrix $R_{ij}(E)$ is always invertible.

In order to calculate the momentum space partial wave separable potential, one needs to evaluate
Eq.~(\ref{eq:2.3.1}),
\begin{equation}
\langle k'|{\bf U}|k \rangle  = \sum_{i,j} \langle k'|u|f_{l,k_{E_i}}\rangle 
\langle f_{l,k_{E_i}} | M | f^*_{l,k_{E_j}}\rangle \langle f^*_{l,k_{E_j}} |u|k\rangle 
\label{eq:2.3.11}
\end{equation}
together with the constraint of Eq.~(\ref{eq:2.3.2}). 
The matrix elements
\begin{equation}
W_{ij} \equiv \langle f_{l,k_{E_i}}| u|f^*_{l,k_{E_j}}\rangle
\label{eq:2.3.12}
\end{equation}
are already part of the computation of the matrix elements $R_{ij}(E)$ from
Eqs.~(\ref{eq:2.3.9}) and (\ref{eq:2.3.10}). The coefficient matrix $M_{ij} \equiv \langle
f^*_{l,k_{E_i}}| M | f_{l,k_{E_j}}\rangle$ is then obtained
as inverse of $W_{ij}$.

%%%%%%%%%%%%%%%%%%%%%%%%%%%%%%%%%%%%%%%%%%%%%%%%%%%%%%%%%%%%%%%%%%%%%%%%%e

\section{Results}

\subsection{On-Shell Behavior of the Optical Potential and its Separable
Representation}
\label{ESTch89}

To demonstrate the construction of a separable representation of
a complex potential we apply the method to n$+^{48}$Ca, n$+^{132}$Sn and n$+^{208}$Pb, and use as a
starting point the Chapel-Hill phenomenological global optical
potential CH89~\cite{Varner:1991zz}. CH89 has been widely used in the literature in
the last decades, and also in recent studies of (d,p) reactions
\cite{Jones:2011kp,Schmitt:2012bt}.  

Using Woods-Saxon functions as parameterized forms, these phenomenological global
optical potentials are most naturally given in coordinate space. For convenience, these global parameterizations assume  local form factors of the interaction but must introduce an explicit energy dependence in the strengths.  

In order to construct a separable momentum space representation of CH89 using the scheme
outlined in the previous section, we first must derive a momentum space
representation of the CH89 optical potential. The explicit evaluation of the Fourier
transform of the CH89 potential is given in Appendix~\ref{appendixA}, including the final expressions we have implemented, 
but there are a few essential points worth highlighting. 
The Fourier transform of the coordinate-space Woods-Saxon function into momentum space can be written as a series
expansion. Fortunately, we found, by the explicit calculation of the leading terms of this expansion,
that only the first two terms are necessary to obtain a converged result. 
These expressions were used as input to a momentum space
LS equation, and the resulting phase shifts were compared to those computed
with the coordinate space CH89 potential using FRESCO~\cite{Thompson:1988zz}. Agreement was found 
for three significant figures.

A separable t-matrix constructed  within the scheme outlined in Section~\ref{formal} 
from a t-matrix calculated as solution of a LS equation with a given potential 
is exact on-shell as well as half-shell at fixed support points $E_i$. For any other
energy $E\neq E_i$ it is then calculated using Eq.~(\ref{eq:2.3.3}).

For studying the quality of the separable representation 
it is convenient to look at the partial wave S-matrix, given as
\begin{equation}
s_l (E) = 1 + 2 i \; {\hat \tau}_l(E), 
\label{eq:3.1.1}
\end{equation}
where $E$ represents the center-of-mass (c.m.) energy, $E\equiv E_{c.m.}$.
The dimensionless amplitude $ {\hat \tau}_l(E)$ is given by
\begin{equation}
{\hat \tau}_l(E) = - \pi \mu k_0 \; t_l(k_0,k_0;E).
\label{eq:3.1.2}
\end{equation}
The on-shell momentum $k_0$ is defined via $E_{c.m.}=k^2_0/2\mu$, and $\mu$ is the reduced mass of the system under consideration. 
The partial wave t-matrix, $t_l(k_0,k_0)$ is either calculated directly from the
CH89 potential or obtained via our general scheme to construct a separable representation thereof. 

\begin{figure}
\begin{center}
\includegraphics[scale=.55]{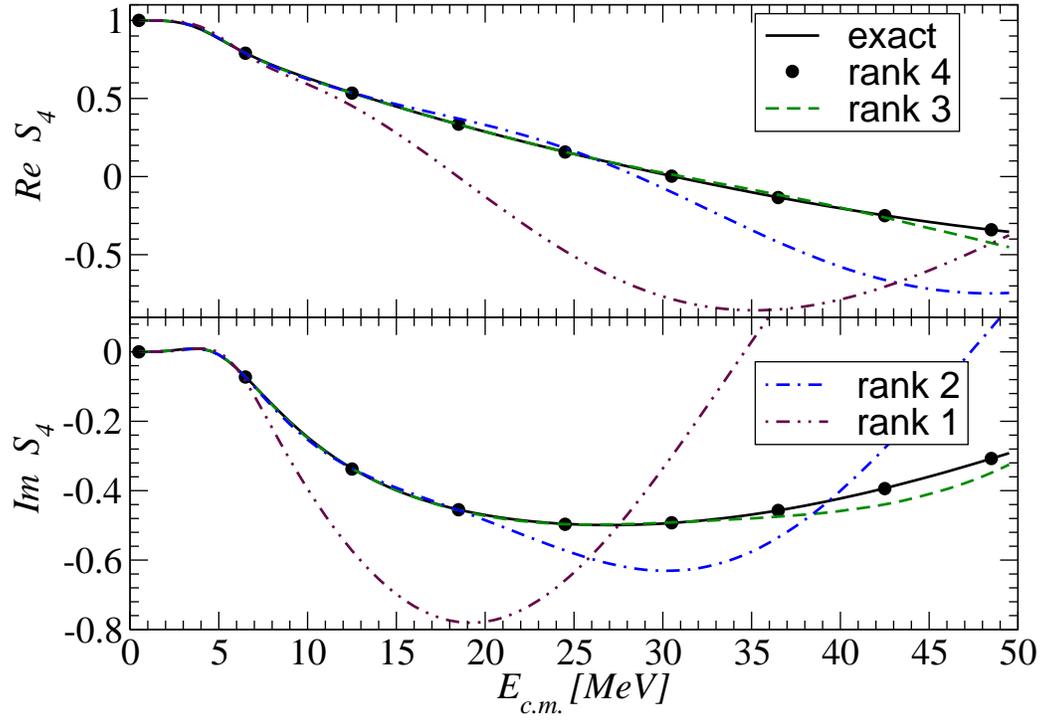}
\vspace{3mm}
\caption{(color online)
The $l=4$ ($j=9/2$) partial wave S-matrix for the n+$^{48}$Ca system obtained from the
CH89~\cite{Varner:1991zz} phenomenological optical potential as function of the c.m energy. 
The exactly calculated S-matrix is given by the solid line. 
Separable representations of rank-1 (support point at 6~MeV),
rank-2 (support points at 6 and 12.5~MeV), and rank-3 (support points at 6, 15, and 25~MeV)
are shown by the dash-double-dotted, dash-dotted,
and dashed lines. The rank-4 representation (support points at 6, 15, 36, and 47~MeV) coincides
with the exact calculation and is indicated by the solid dots.
\label{fig1}
}
\end{center}
\end{figure}

First we study in detail the scheme for constructing a separable representation of 
the n+$^{48}$Ca system for c.m.
energies from 0 to 50~MeV. Our goal is to arrive at an excellent separable representation of the
partial wave S-matrix starting from the local CH89 potential. 
For practical applications in e.g. three-body type calculations, it is desirable to 
achieve this with as low a rank as possible.  As a representative case, we show the
$l=4$, $j=9/2$ partial wave S-matrix in Fig.~\ref{fig1}. The S-matrix obtained from the 
solution of the LS
equation with the original CH89 optical potential (solid line) shows a relatively mild variation with
energy in the energy regime under consideration. If one is only interested in describing
the very low energies, i.e. $E_{c.m.} \leq 10$~MeV, a rank-1 separable potential 
with a support point at 6~MeV is barely
sufficient (dash-double-dotted line), while a rank-2 representation with support points at 6 and
12.5~MeV can already capture the range between 0 and 20~MeV relatively well. A rank-3
representation with support points at 5, 15, and 25~MeV captures the S-matrix up to roughly
35~MeV. However, for a high quality separable representation of at least 4  significant figures of
the CH89 result  a rank-4 representation with support points at 6, 15, 36,
and 47~MeV is needed in this partial wave. 
The figure also shows that more support points are needed in the region where the S-matrix 
shows structure, and less points are necessary for the smooth region. 

\begin{figure}
\begin{center}
\includegraphics[scale=.55]{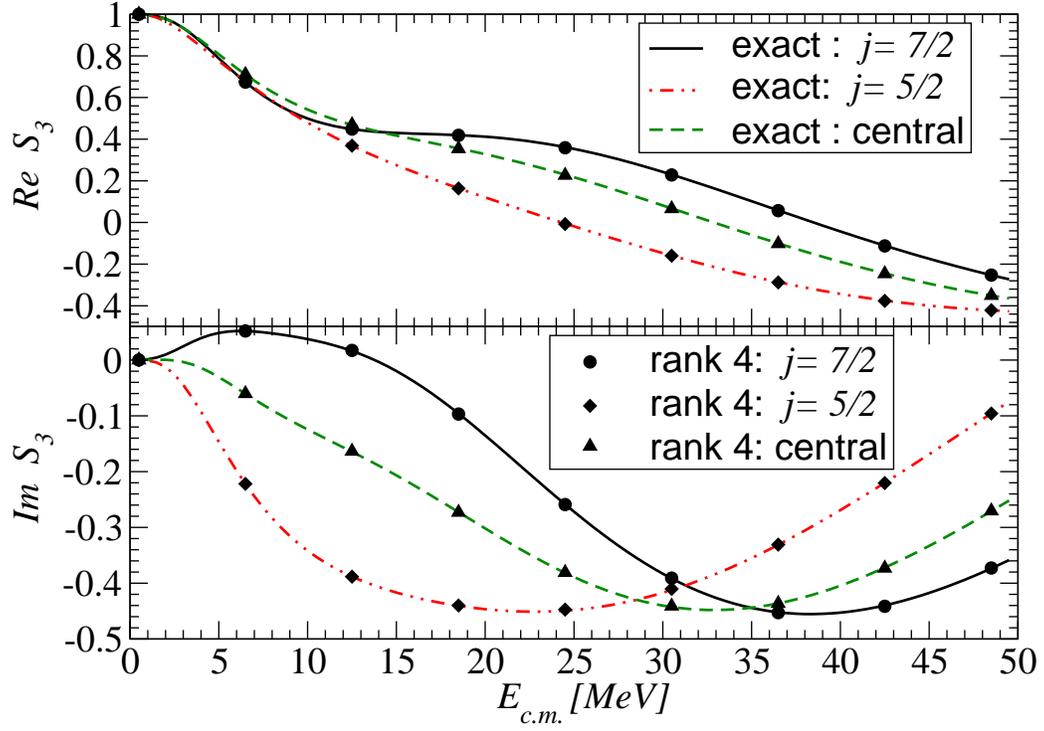}
\vspace{3mm}
\caption{(color online) 
The $l=3$  partial wave S-matrix for the n+$^{48}$Ca system obtained from the
CH89~\cite{Varner:1991zz} phenomenological optical potential as function of the c.m energy.
The exactly calculated partial wave S-matrices for $j=7/2$, $j=5/2$, and the central part of the
optical potential alone are given by the solid, dash-double-dotted and dashed lines. The results
for the corresponding rank-4 separable representations (support points at  6, 15, 36, and 47~MeV)
are overlayed and indicated by the solid symbols as indicated in the figure.
\label{fig2}
}
\end{center}
\end{figure}

The next question for a practical implementation of the EST scheme is whether the optimum 
support points when including both central and spin-orbit interactions differ from the results when including the central interaction only. 
Usually the central part of an optical potential is larger than the spin-part. Thus one may
expect it to be sufficient to find EST support points for the S-matrix computed using 
only the central part of the optical potential, and then use the same points for 
deriving the separable representation of the $l\pm 1/2$ partial wave S-matrices.
This is indeed the case, as is demonstrated for the $l=3$ partial wave S-matrix for 
the n+$^{48}$Ca system in Fig.~\ref{fig2}. 
The dashed line shows the S-matrix calculated from the cental part of the CH89 optical potential, 
while the solid and dash-double-dotted lines represent the $l+1/2$ and the $l-1/2$ partial 
wave S-matrices. Our results show that it is indeed sufficient to determine the
EST support points for the S-matrix computed from the cental part of the optical potential.  Thus, once the support points are determined from a calculation including the central part, one only needs to replace the corresponding form factors, i.e. the half-shell t-matrices, at the support points with the ones containing the spin-orbit contribution to obtain the separable representation of the $l\pm 1/2$ partial 
wave S-matrices with unchanged accuracy. 

\begin{figure}
\begin{center}
\includegraphics[scale=.55]{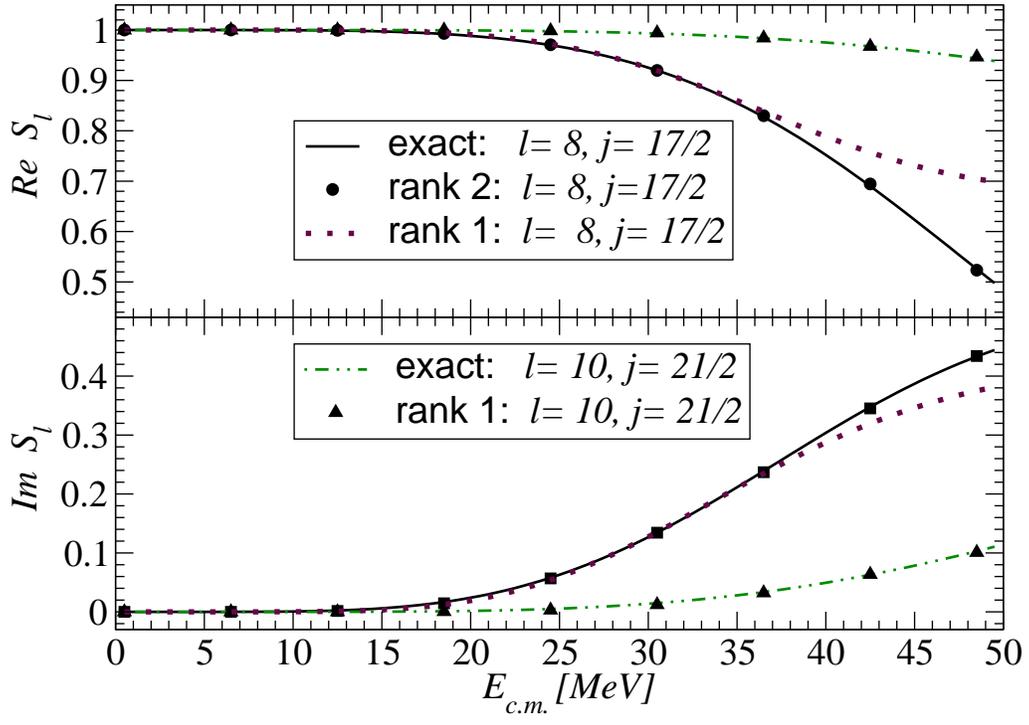}
\vspace{3mm}
\caption{(color online) 
The $l=8, j=17/2$ (solid line) and $l=10, j=21/2$  (dash-double-dotted line)
partial wave S-matrices for the n+$^{48}$Ca system calculated from the
CH89~\cite{Varner:1991zz} phenomenological optical potential as function of the c.m energy.
The rank-1 (support point at 29~MeV) and rank-2 (support points at 29 and 47~MeV) representations
for the $l=8$ partial wave S-matrix are given by the dotted line and the filled circles.
The rank-1 (support point at 40~MeV) representation for the $l=10$ partial wave S-matrix is
indicated by the filled triangles.
\label{fig3}
}
\end{center}
\end{figure}

For the n+$^{48}$Ca system we find that for the lower partial waves a rank-4 separable
representation is sufficient for energies up to $E_{c.m.}$~=~50~MeV. Let us now consider what
happens as we increase the angular momentum. Reaction calculations in the energy range $0-50$ MeV
often require partial waves up to $l=20$, but due to the centrifugal barrier with increasing
angular momentum the t-matrix remains close to 
zero even  at higher scattering energy. Therefore, one may expect, for the same accuracy, 
a lower rank separable representation to suffice for representing higher partial waves. 
This is indeed the case as demonstrated in Fig.~\ref{fig3}. Here we show the $l=8$, $j=17/2$ and $l=10$, $j=21/2$ partial wave S-matrices and their corresponding separable representation. 
We find that for $l=10$ a rank-1 representation is sufficient up to $E_{c.m.}$~=~50~MeV, whereas the $l=8$ partial wave S-matrix still requires a rank-2 representation.  
For the energy regime under consideration, we determined the angular momentum intervals in which 
separable representations of specific ranks represent the original partial wave S-matrices 
within 4 significant figures. 
Those angular momentum groups and their corresponding support points are listed in Table~\ref{table-4}.

\begin{figure}
\begin{center}
\includegraphics[scale=.55]{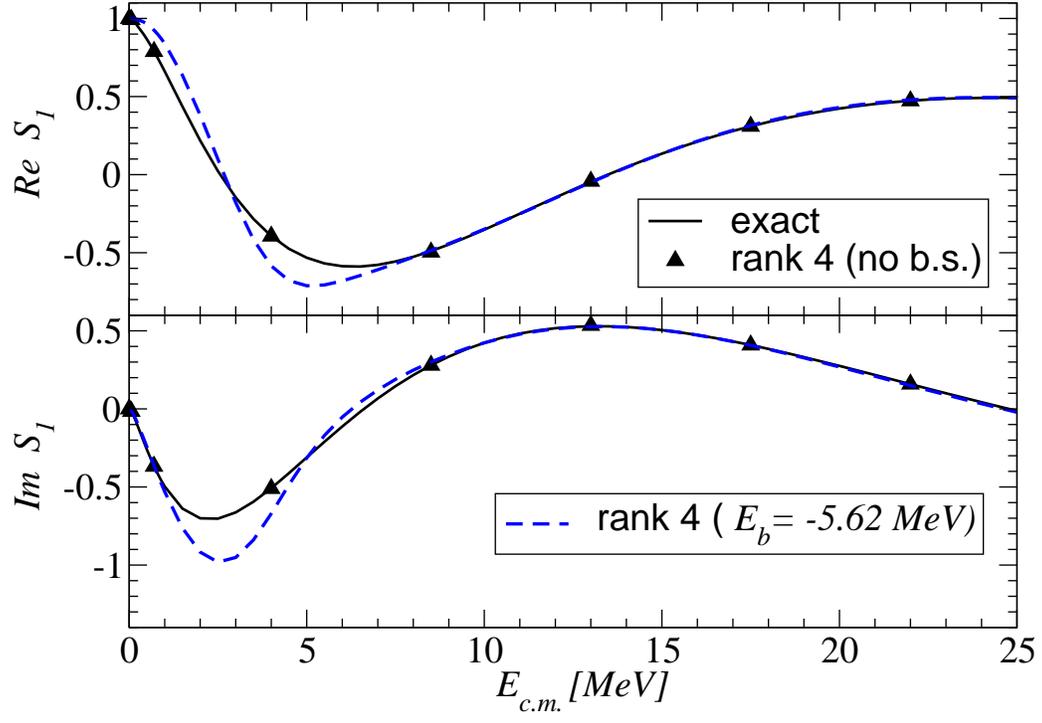}
\vspace{3mm}
\caption{(color online) 
The $l=1$ ($j=3/2$) partial wave S-matrix for the n+$^{48}$Ca system obtained from the
CH89~\cite{Varner:1991zz} phenomenological optical potential as function of the c.m energy.
The exactly calculated S-matrix is given by the solid line. 
A separable representation of rank-4 (support points at  6, 15, 36, and 47~MeV) is indicated by
the filled triangles. For the dashed line the lowest support point at 6~MeV is replaced by a
point at the bound state $E_b$~=~-5.62~MeV. 
\label{fig4}
}
\end{center}
\end{figure}

Support points need not be at positive energies, therefore a bound state may
be included and this is exactly  the situation in the so-called unitary pole
approximation~\cite{Lovelace:1964mq}. 
Thinking about our original purpose (namely  Faddeev description of (d,p) reactions), a good
description of the final bound state will be necessary. To understand the effect of including a
bound state in the separable representation, we consider the $l=1$, $j=3/2$ partial wave S-matrix
of the n+$^{48}$Ca system, which would correspond to the partial wave of the valence neutron in
the ground state of $^{49}$Ca. When extrapolating the CH89 parameterization to $E_{c.m.}<0$, 
we set the imaginary part to zero and directly extrapolate the real part with no further fitting, 
just for the sake of illustrating the method here. With such an approach, we find two
bound states with energies $E_b$~=~-28.8~MeV and $E_b$~=~-5.62~MeV. We now move the
support point at 6~MeV to $E_b$~=~-5.62~MeV, which would be the valence orbital in this
system. The results are shown in Fig.~\ref{fig4}. The solid line shows the exact
calculation, the solid triangles its separable rank-4 representation.  The so obtained
rank-4 separable representation is given by the dashed line in Fig.~\ref{fig4}. Though the bound state is relatively close to threshold, the deviation from the original representation and the exact S-matrix is quite large in the
region between 1 and 7~MeV, indicating that the low energy support point is needed for capturing the structure of this partial wave S-matrix at low energies. 

From this study we learn that
though it is very easy to add bound states to the separable representation of a specific partial wave 
S-matrix, such a bound state, even if shallow, does not necessarily replace a low energy support point. For an application to (d,p) reactions, one should add a  bound state corresponding to the final state populated through the reaction, thus increasing the rank by 1.

Up to now we studied the  separable representations in detail
 for the n+$^{48}$Ca system. Next, we turn to a heavy nuclei and
repeat the study for  the n+$^{132}$Sn  and n+$^{208}$Pb systems. 
In these cases, the partial wave S-matrices have  more structure, 
and thus require more  support points for an accurate representation. 
For different partial waves, we again find that we have groups of angular momenta for which a specific 
rank is required to represent the partial wave S-matrix to an accuracy of at least 4 significant figures.
These groups and the energies of the corresponding support points are summarized in 
Table~\ref{table-4}. First we note that a separable representation 
of rank-5 is sufficient for the low angular momentum states. Similar to the n+$^{48}$Ca system, 
the higher angular momentum states require successively less support points. 
However, in order for a rank-1 representation to be accurate in the energy regime 
under consideration, one must go up to $l=16$. 

Considering the number of support points needed for medium-mass (n+$^{48}$Ca) and heavy
( n+$^{132}$Sn  and n+$^{208}$Pb) systems, we find it encouraging that a very good description of the heavy
systems can be achieved by increasing the rank of the representation by only one, relative to
the medium-mass case. 

\begin{figure}
\begin{center}
\includegraphics[scale=.55]{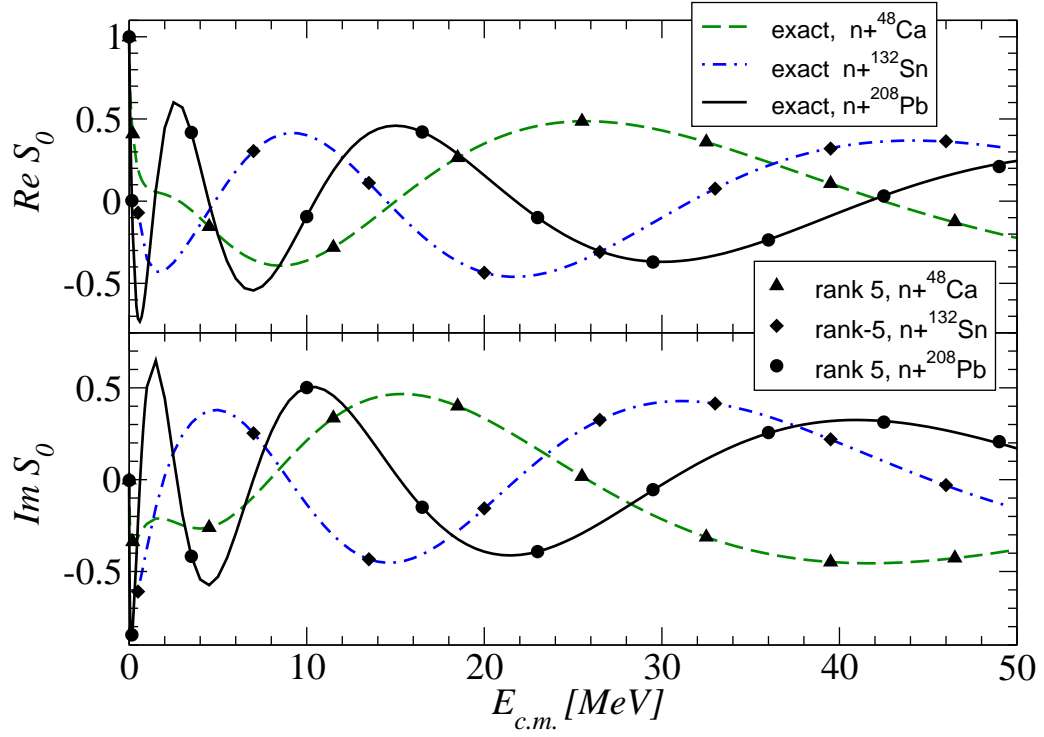}
\vspace{3mm}
\caption{
The $l=0$ partial wave S-matrices for the n+$^{48}$Ca system (dashed line), the n+$^{132}$Sn
(dash-dotted line),
and for the  n+$^{208}$Pb system (solid line)  calculated exactly from the
CH89~\cite{Varner:1991zz} phenomenological optical potential as function of the c.m energy.
A rank-5 representation (support points at 5, 11, 21, 36, and 47~MeV) for both systems is
overlayed and indicated by the filled circles (n+$^{208}$Pb), the filled diamonds (n+$^{132}$Sn),
 and filled triangles (n+$^{48}$Ca). 
\label{fig5}
}
\end{center}
\end{figure}

If we are interested in a `universal' separable representation for partial waves t-matrices 
obtained from the CH89 optical potential, we can use the  support points obtained for 
the n+$^{208}$Pb system, apply them to all other systems and obtain a high quality 
separable representation. This is shown in Fig.~\ref{fig5} for the $l=0$ partial wave S-matrix 
for the n+$^{48}$Ca, n+$^{132}$Sn and n+$^{208}$Pb systems. The  points used here can also be used for 
the relevant higher angular momentum t-matrices.

\subsection{Off-Shell Behavior of the Optical Potential and its Separable
Representation}
\label{offshell}

After having established the scheme for finding separable representations of 
the t-matrices
obtained from the CH89 phenomenological optical potential and checking its accuracy by comparing
the S-matrix elements between 0 and 50~MeV, we now want to look at the resulting off-shell
t-matrices. 
The partial wave off-shell t-matrices, $t_l(k',k;E)$, are calculated at a given energy
$E$  using Eq.~(\ref{eq:2.3.3}) sandwiched with arbitrary moment $k'$ and $k$.

\begin{figure}
\begin{center}
\includegraphics[scale=1.50]{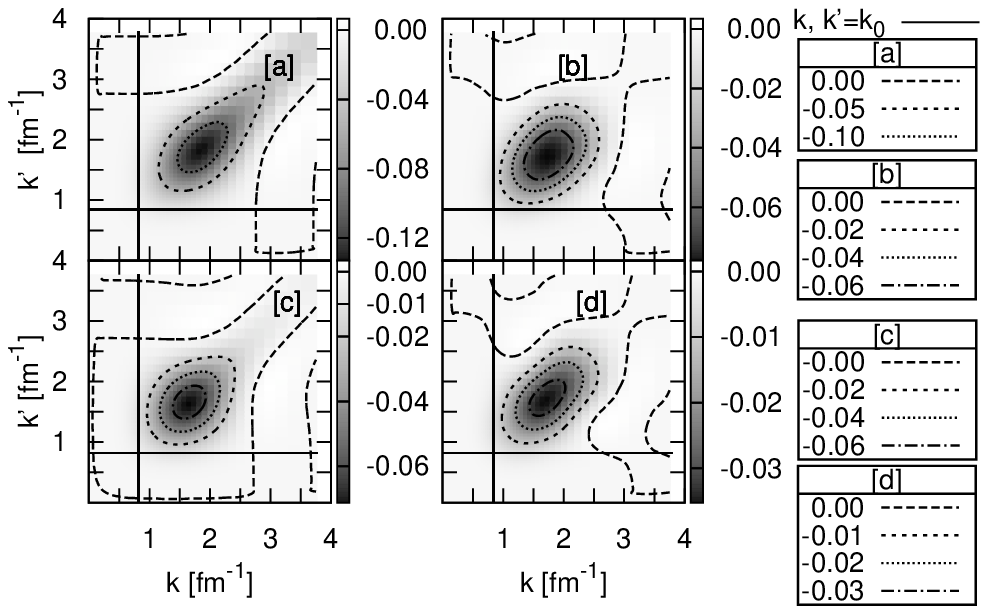}
\vspace{3mm}
\caption{
The $l=6$, $j=13/2$ partial wave off-shell t-matrix elements, $t_6(k',k;E_{c.m.})$,
 for the n+$^{48}$Ca system 
computed at $E_{c.m.}$~=~16~MeV as function of the off-shell momenta $k'$ and $k$. 
The on-shell momentum, $k_0$~=~0.87~fm$^{-1}$, is indicated by the straight lines. 
Panels [a] and [c] on the left hand side
show the real and imaginary part of the t-matrix in units fm$^2$ 
obtained from the
CH89~\cite{Varner:1991zz} phenomenological optical potential,
while panels [b] and [d] on the right hand side
 depict their separable representation (rank-3). Note the difference
in scale between the left and right hand side panels.
\label{fig6}
}
\end{center}
\end{figure}

In Fig.~\ref{fig6} we show the off-shell t-matrix,  at $E_{c.m.}$~=~36~MeV 
for the n+$^{48}$Ca system in the partial wave $l=6$, $j=13/2$. The left side panels ([a] and [c])
show the off-shell t-matrix as function of momenta $k$ and $k'$ 
obtained as solution of the LS equation with the CH89 potential, 
while the right side panels ([b] and [d]) depicts their separable representation. The on-shell
momentum at $k_0$~=~1.3~fm$^{-1}$ is indicated in each panel by the horizontal/vertical lines. 

First, we point out that in both cases the t-matrix is symmetric around the line $k'=k$,
which must be the case if the underlying potential fulfills the reciprocity theorem. This also
shows that our generalization of the EST scheme to complex potentials is correct.
Had we used the definition given in Ref.~\cite{Ernst:1973zzb}, the 
resulting off-shell t-matrix would not have this symmetry.

Second, we observe that overall the magnitude of the off-shell elements of the separable interaction is smaller  than that of the t-matrix of the original potential.
It was shown in Ref.~\cite{Ernst:1974zzb} that, for  a rank-1 separable potential, the off-shell t-matrix has the form $v(p)/v(k_0)$, where $k_0$ is the on-shell momentum, the form factors
$v(p)$ are the half-shell t-matrices calculated at the support points and  $v(k_0)$
their on-shell value. 
This justifies why the magnitude of the off-shell elements of the separable
t-matrix is smaller than the magnitude of those obtained from the solution of an LS equation with the
original CH89 potential. 

Furthermore, we note that the t-matrix obtained from the CH89 potential has significant 
non-vanishing values along the line $k=k'$ even for momenta $k=k' \ge 4$~fm$^{-1}$, which is typical for local potentials,
while these off-shell matrix elements approach zero for the separable counterpart.
The fact that the separable representation projects out high momentum components of the original
t-matrix is reminiscent of renormalization group techniques~\cite{Bogner:2006pc}, which in nuclear physics are typically applied to the NN force. We note that the off-shell dip around $k'=k \sim$~1.5~fm$^{-1}$ is present in both cases, 
indicating that closer to the on-shell point, the EST scheme preserves the off-shell structure. 

\begin{figure}
\begin{center}
\includegraphics[scale=1.50]{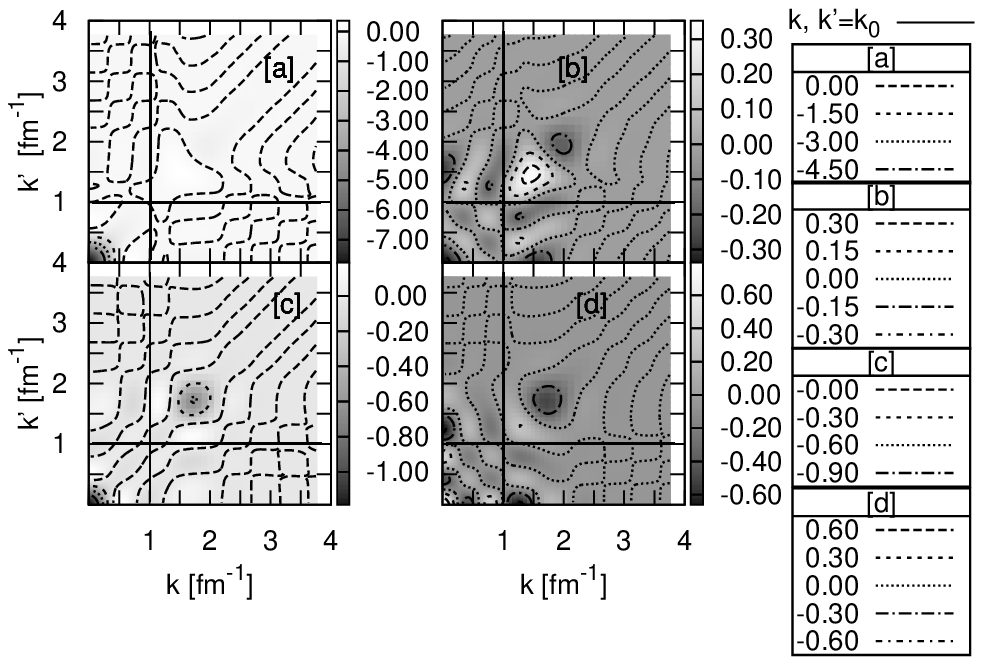}
\vspace{3mm}
\caption{
The $l=0$ partial wave off-shell t-matrix elements, $t_0(k',k;E_{c.m.})$, 
for the n+$^{208}$Pb system
computed at $E_{c.m.}$~=~21~MeV as function of the off-shell momenta $k'$ and $k$.
The on-shell momentum, $k_0$~=~1.0~fm$^{-1}$, is indicated by the straight lines.
Panels [a] and [c] on the left hand side 
show the real and imaginary part of the t-matrix in units fm$^2$ 
obtained from the
CH89~\cite{Varner:1991zz} phenomenological optical potential,
while panels [b] and [d] on the right hand side depict their separable representation (rank-3).
Note the difference in scale between the left and right hand side panels.
\label{fig7}
}
\end{center}
\end{figure}

In Fig.~\ref{fig7} we show the off-shell t-matrix at $E_{c.m.}$~=~21~MeV for the  n+$^{208}$Pb
system in the partial wave $l=0$. The left panel again depicts the real and imaginary t-matrix
obtained as solution of the LS equation with the CH89 potential, while the right side panels
give their separable representation. The on-shell momentum $k_0$~=~1.0~fm$^{-1}$ is indicated by 
a horizontal/vertical lines.  The scale of the exact solution is dominated by the strong dip close to the
origin. This dip is also present in the separable representation, however almost a factor of four
smaller in case of the real part and a factor of two for the imaginary part.  
The off-shell structure around $k'=k \sim$~1.5~fm$^{-1}$ is again captured  well by the
separable representation. The exact solution also has non-zero contributions along the line
$k'=k$, while the separable representation does not, however due to the scale this is not visible
in the figure.  For these heavier systems, the higher partial waves behave very similar
 to the ones in the n+$^{48}$Ca
system. Therefore we do not show them separately.

\section{Summary and Conclusions}
\label{summary}

In this work we extended the well-known EST scheme~\cite{Ernst:1973zzb} for creating separable
representations of two-body transition matrix elements as well as potentials to the realm of
complex potentials. Requiring that the separable transition matrix fulfill the reciprocity
theorem, we identified a suitable rank-1 separable potential. In analogy to
Ref.~\cite{Ernst:1973zzb}, we generalized this potential to arbitrary rank.

All calculations presented in this work are based on the Chapel Hill phenomenological optical
potential CH89~\cite{Varner:1991zz}. Since the CH89 potential, as nearly all phenomenological
optical potentials, is given in coordinate space using Woods-Saxon functions, we first 
give a semi-analytic Fourier transform of those Woods-Saxon functions in terms of a series
expansion. In practice, it turns out that only two terms in the
expansion are sufficient for achieving convergence. Note that our approach for deriving the momentum-space optical potential is general and can be applied to any optical potential of Woods-Saxon form.
This momentum space CH89 potential is then used in the partial-wave LS integral equation to
calculate  half-shell t-matrices. These then serve as input to the 
the generalized scheme for creating separable representations for complex potentials.

The systematic studies in this paper include: n+$^{48}$Ca, n+$^{132}$Sn and n+$^{208}$Pb. 
We are able to provide, for all cases, a systematic classification of support points for partial-wave groups, so that
the partial-wave S-matrices are reproduced to at least 4 significant figures compared to the
original momentum space solution of the LS equation. We find the low partial waves of the
n+$^{208}$Pb system require a rank-5 separable potential to be well represented in the energy
regime between 0 and 50~MeV center-of-mass energy. The support points obtained for this case are
well suited to represent all partial waves of the n+$^{208}$Pb as well as all lighter systems
described by the CH89 optical potential.

We find that the rank required for achieving a good representation decreases with increasing
angular momentum of the partial wave considered. We developed recommendations for both the rank
and the locations of support points to be used when describing medium-mass and heavy systems
generated from the CH89 potential.  Our recommendations group together partial waves. We also demonstrated that it is sufficient to determine support points including only the central part of the
optical potential; when the spin-orbit interaction is added and the form factors are accordingly modified, the same support
points can be expected to yield  a good representation.

We then investigated the off-shell behavior of the constructed separable representations, and
found that overall, the  high momentum components along the $k=k'$ axis which are typical for
local potentials are removed from the separable representation. Furthermore, the off-shell
elements of the separable representation are smaller in magnitude, however follow the functional
shape of the CH89 potential. Since off-shell matrix elements are not observables, only reaction
calculations can show if the differences  seen in the off-shell t-matrix have 
any consequences for e.g. three-body observables.  Future
work will address this question.

%-------------------------------------------------------------------------------
%****************************************************************************

%\newpage

\appendix

\section{Fourier Transform of  Global Optical Potentials}
\label{appendixA}

\subsection{Global Optical Potentials in Coordinate Space}

Most global optical potentials are parameterized in terms of Wood-Saxon form factors and its derivatives.
For convenience, here we provide the definitions of the form factor and standard parameterizations of 
phenomenological optical potentials. These definitions are also those for the CH89 potential~\cite{Varner:1991zz}, which we use throughout this work.
The short range, nuclear part is given by
\begin{equation}
U_{nucl}(r)= V(r)+i\big[ W(r)+W_s(r)\big]+V_{ls}(r)\; {\bf l}\cdot {\bsigma},
\label{eq:a.1.1}
\end{equation}
where
\begin{eqnarray}
V(r) &=&  -V_r f_{ws}(r,R_0,a_0) \cr
W(r) &=& -W_v f_{ws}(r,R_s,a_s) \cr
 W_s(r)  &=& -W_s(-4a_s)f_{ws}'(r,R_s,a_s) \cr
V_{ls}(r) &=& -(V_{so}+iW_{so})(-2)g_{ws}(r,R_{so},a_{so}),
\label{eq:a.1.2}
\end{eqnarray}
\begin{eqnarray}
f_{ws}(r,R,a) &=& \frac{1}{1+\exp(\frac{r-R}{a})} \cr 
f_{ws}'(r,R,a) &=& \frac{d}{dr}f_{ws}(r,R,a)\cr 
g_{ws}(r,R,a) &=& f'_{ws}(r,R,a)/r \; .  
\label{eq:a.1.3} 
\end{eqnarray}
The constants $V_r$, $W_v$, $V_{so}$, and $W_{so}$ are the strength parameters, and
$a$ and $R$ the diffuseness and the radius parameters given in
Ref.~\cite{Varner:1991zz}.

\subsection{Fourier Transform of Woods-Saxon Functions}

The basic functions to be transformed to momentum space are given by
Eqs.~(\ref{eq:a.1.3}), which are the Woods-Saxon function and its derivative. The
Fourier transform of e.g. $V(r)$ of Eq.~(\ref{eq:a.1.2}) is given by
\begin{equation}
V(q)=\frac{1}{2\pi^2}\frac{1}{q} \int_0^\infty dr \; r V(r) \sin(qr),
\label{eq:a.2.1}
\end{equation}
where $q$ is the momentum transfer defined as $q=|{\bf q}|= |{\bf k'}-{\bf k}|$.

Introducing dimensionless variables $\rho_k= qa_k$, $z=r/a_k$, $\alpha_k= R_k/a_k$ and 
$\gamma_k= e^{-\alpha_k}$ (with $k\equiv 0,s,so$), and inserting into  Eq.~(\ref{eq:a.2.1})  yields
\begin{equation}
V(q)=-\frac{V_r}{2\pi^2}\frac{a_0^2}{q} \int_0^\infty dz \; \frac{z \sin (\rho_0
z)}{1+\exp(z-\alpha_0)}\; \equiv
\; -\frac{V_r}{2\pi^2}\frac{a_0^2}{q}\; \Im m \;\int_0^\infty dz \; \frac{z \exp (i\rho_0 z)}{1+\exp(z-\alpha_0)}.
\label{eq:a.2.2a}
\end{equation}

Repeating the above step for all the expressions in Eqs.~(\ref{eq:a.1.2}) yields
expressions that are similar to Eq.~(\ref{eq:a.2.2a}). The  are four distinct  real integrals appearing in
the Fourier transform expressions of Eqs.~(\ref{eq:a.1.2}) and they are the real and imaginary parts of 
the two complex integrals  $I_1$ and $I_2$ such that
\begin{eqnarray}
I_1(\rho_k,\alpha_k)&=& \int_0^\infty dz \frac{z\exp (i\rho_k z)}{1+\exp(z-\alpha_k)}
\equiv \int_0^\infty f_1(z) dz, \cr
I_2(\rho_k,\alpha_k) &=& \int_0^\infty dz \frac{\exp (i\rho_k z)}{1+\exp(z-\alpha_k)}
\equiv \int_0^\infty f_2(z) dz,
\label{eq:a.2.2}
\end{eqnarray}
 where $\rho_k$ and $\alpha_k$ are real numbers. 
With the definitions  given in Eq.~(\ref{eq:a.2.2}) the Fourier transforms of  
the expressions given in Eqs.~(\ref{eq:a.1.2}) can be written as
\begin{eqnarray}
 V(q)&=& -\frac{V_r}{2\pi^2} \frac{a_0^2}{q}\; \Im m \; I_1(\rho_0,\alpha_0) \cr
W(q)&=&  -\frac{W_v}{2\pi^2} \frac{a_s^2}{q}\; \Im m\; I_1(\rho_s,\alpha_s) \cr
W_s(q)&=& -\frac{2W_s}{\pi^2}  a_s^2 \left[\frac{1}{q}\;\Im m\; I_2(\rho_s,\alpha_s)+a_s\; \Re e \;I_1(\rho_w,\alpha_s)\right] \cr
 V_{ls}(q)&=&- \frac{(V_{so}+W_{so})}{\pi^2} a_{so}\; \Re e \;
I_2(\rho_{so},\alpha_{so})\; .
\label{eq:a.2.3}
\end{eqnarray}
The integrals $I_1$ and $I_2$ can be evaluated by contour integration. 
The closed integration loop is taken to be the boundary of the first quadrant 
(i.e from $0$ to $\infty$ along positive real axis, then a circular path from $+\infty$ 
to $+i\infty$, and finally from $+i\infty$ to $0$ along the positive imaginary axis). 
The contribution from the circular path integral is zero in this case.
Both integrands have poles at $z_n= \alpha+i\pi(2n+1)$ with $n\in \mathbf Z$. 
The corresponding residues are given by
\begin{eqnarray}
Res(z_n,f_1) &=& -[\alpha+i\pi(2n+1)]e^{i\rho\alpha}e^{-(2n+1)\pi\rho} \cr
 Res(z_n,f_2) &=& -e^{i\rho\alpha}e^{-(2n+1)\pi\rho}\; .
\label{eq:a.2.4}
\end{eqnarray}
The total closed loop integral is obtained via the residue theorem, and the path from
$+i\infty$ to $0$ is evaluated as series expansion of the integrand in terms of
$\gamma$. This leads to
 \begin{eqnarray}
\Re e\; I_1(\rho,\alpha)&=& \frac{2\pi e^{-\pi \rho}}{(1-e^{-2\pi\rho})^2}\left[\pi (1+e^{-2\pi \rho})\cos (\rho \alpha) 
  +\alpha(1-e^{-2\pi\rho})\sin (\rho \alpha)\right] \cr
&& -\sum_{n=0}^\infty (-1)^n\gamma^n\frac{\rho^2-n^2}{(\rho^2+n^2)^2} \cr
\Im m\; I_1(\rho,\alpha) &=& \frac{2\pi e^{-\pi \rho}}{(1-e^{-2\pi\rho})^2}\left[\pi (1+e^{-2\pi \rho})\sin (\rho \alpha) 
  -\alpha(1-e^{-2\pi\rho})\cos (\rho \alpha)\right] \cr
 && -2\sum_{n=0}^\infty (-1)^n\gamma^n\frac{n\rho}{(\rho^2+n^2)^2} \cr
\Re e\; I_2(\rho,\alpha) &=& \frac{2\pi e^{-\pi \rho}}{1-e^{-2\pi\rho}}\sin (\rho \alpha) 
  -\sum_{n=0}^\infty (-1)^n \gamma^n \frac{n}{\rho^2+n^2} \cr
\Im m \; I_2(\rho,\alpha)&=& -\frac{2\pi e^{-\pi \rho}}{1-e^{-2\pi\rho}}\cos (\rho \alpha) 
  +\sum_{n=0}^\infty (-1)^n \gamma^n \frac{\rho}{\rho^2+n^2} \; .
\label{eq:a.2.5}
\end{eqnarray}
Similar expressions were derived in Ref.~\cite{RodriguezGallardo:2007dc}.
The infinite sums can not be evaluated analytically and their convergence 
properties need to be studied numerically. 
In order to do so, let us define the finite 
sums $S_1^R$, $S_1^I$, $S_2^R$, and $S_2^I$ as
\begin{eqnarray}
S_1^R &=& -\sum_{n=0}^{n_{max}} (-1)^n\gamma^n\frac{\rho^2-n^2}{(\rho^2+n^2)^2} \cr
S_1^I &=& -2\sum_{n=0}^{n_{max}} (-1)^n\gamma^n\frac{n\rho}{(\rho^2+n^2)^2} \cr
S_2^R &=& -\sum_{n=0}^{n_{max}} (-1)^n \gamma^n \frac{n}{\rho^2+n^2} \cr
S_2^I &=& \sum_{n=0}^{n_{max}} (-1)^n \gamma^n \frac{\rho}{\rho^2+n^2} \; .
\label{eq:a.2.6}
\end{eqnarray}
The convergence of the above sums is studied as function of $n_{max}$.
The parameter crucial to the convergence  is $\gamma = e^{-\alpha}$,
the smaller $\gamma$, the faster the series will converge. 
Thus, the slowest convergence will be for the smallest values of $\alpha$ for a 
given optical potential.
For the CH89 optical potential~\cite{Varner:1991zz} the minimum value is slightly 
larger than 5.8 corresponding to mass number $A$~=~40. For all heavier nuclei $\alpha$ 
is larger. The numerical values of the sums of Eqs.~(\ref{eq:a.2.6}) are given in 
Table~\ref{table-2} as function of $n_{max}$ for the worst case of $\alpha$~=~5.8. 
The Table demonstrates that for the CH89 global optical potential $n_{max}$~=~2 
is sufficient to obtain convergence up to six decimal points.

The truncation discussed above focused on CH89. The conditions for convergence should be tested for other global optical potentials. We repeated the study for the Weppner-Penny~\cite{Weppner:2009qy} (WP) global optical potential, which was fitted to nuclei
as light as $^{12}$C. Then, the required minimal value of 
$\alpha$ is about 3, where we consider that  $\alpha$ is energy dependent. 
In Table~\ref{table-3},
we present the convergence of the sums of  Eqs.~(\ref{eq:a.2.6}) 
for the fixed value $\alpha$~=~3.0, when using the WP potential.  We find that $n_{max}$~=~4 is sufficient to obtain convergence up to six decimal points. For the heavier nuclei, $A \geq 40$, the WP optical potential requires only
$n_{max}$~=~2 similar to the CH89 optical potential.

\subsection{The Partial Wave  Optical Potential in Momentum Space}

According to the considerations in the previous section the first two terms in the
series expansion are sufficient to obtain  momentum space expressions for the
Fourier transforms of the Woods-Saxon functions of the CH89 potential which
are sufficiently accurate as shown in Table~\ref{table-2}. This leads
to
\begin{eqnarray}
{V} ( q)& =& \frac{V_r}{\pi^2}
 \Bigg\{ \frac{\pi a_0 e^{-\pi a_0 q}} 
{q\left(1-e^{-2\pi a_0 q}\right)^2} \left[R_0 \left(1-e^{-2\pi a_0 q}\right)
\cos\left(q R_0\right) -
 \pi a_0 \left(1+e^{-2\pi a_0 q}\right) \sin\left(q R_0\right)\right] \cr
& & - a_0^3 e^{-\frac{R_0}{a_0 }}  \left[ \frac{1}{(1+a_0 ^2q^2)^2} - 
\frac{2 e^{-\frac{R_0}{a_0 }}} {(4+a_0 ^2q^2)^2}\right] \Bigg\} \; 
\label{eq:a.3.1}
\end{eqnarray}
for the Fourier  transform of $V(r)$. Here  $q$ stands for the
magnitude of the momentum transfer as defined in the previous section. 

For ${W}( q)$, the imaginary volume term, the same
expression is obtained with $W_v$, $a_s$, and $R_s$ in the place of
$V_r$, $a_0$, and $R_0$.

For the surface term $W_s (r)$ of  Eq.~(\ref{eq:a.1.2}) we obtain the
following momentum space form
\begin{eqnarray}
{W}_s (q)& =& - 4 a_s \;\frac{W_s}{\pi^2} \;
 \Bigg\{ \frac{\pi a_s e^{-\pi a_s q}}{\left(1-e^{-2\pi a_s q}\right)^2} \cr
& \Big[& \Big( \pi a_s \left(1+e^{-2\pi a_s q}\right) -\frac{1}{q}
\left(1-e^{-2\pi a_s q}\right) \Big) \cos(qR_s) + R_s \left(1-e^{-2\pi
a_s q}\right) \sin(q R_s) \Big]  \cr
&+& a^2 e^{-R_s/a_s } \left[ \frac{1}{(1+a_s ^2 q^2)^2} - \frac{4
e^{-R_s/a_s }}{(4+a_s ^2 q^2)^2} \right] \Bigg\} \; .
\label{eq:a.3.2}
\end{eqnarray}
The Fourier transform of the spin-orbit term, $V_{ls}(r)$, leads to
\begin{equation}
{V}_{ls}(q) = -\frac{a_{so}}{\pi^2}(V_{so}+iW_{so})\Bigg\{ \frac{2\pi
e^{-\pi
a_{so}q}}{1-e^{-2\pi a_{so}q}}
\sin(qR_{so})+e^{\frac{-R_{so}}{a_{so}}}\left(\frac{1}{1+a_{so}^2q^2}-\frac{2e^{-\frac
{R_{so}}{a_{so}}}}{4+a_{so}^2q^2}
\right)\Bigg\}.
\label{eq:a.3.3}
\end{equation}
%Since $q$ is a function of ${\bf k}$ and ${\bf k'}$, the potential terms $V(q)$, $W(q)$, $W_s(q)$,
%and $V_{ls}(q)$ are often written as  $V({\bf k^\prime},{\bf k})$, $W({\bf k^\prime},{\bf k})$, $W_s({\bf k^\prime},{\bf k})$,
%and $V_{ls}({\bf k^\prime},{\bf k})$ in the momentum space representation.
Here ${\bf q}$ is the momentum transfer, ${\bf q} = {\bf k}' -{\bf k}$ with 
$|{\bf q}| = \sqrt{k'^2 + k^2 - 2 k'k\cos \theta_{k'k}}$. 

The most general
form of a momentum space potential is usually written as $\langle {\bf k}'| U_{nucl}| {\bf
k}\rangle \equiv  U_{nucl}({\bf k}',{\bf k})$, with ${\bf k'}$ and ${\bf k}$ being linear
independent vectors spanning the scattering plane. A local potential only depends on the
momentum transfer, ${\bf k}' -{\bf k}$, and has no contribution proportional to the orthogonal 
vector $\frac{1}{2} ({\bf k}' +{\bf k})$.
Thus, in momentum space the most general form of an optical potential of Eq.~(\ref{eq:a.1.1}) 
takes the form
\begin{equation}
 U_{nucl}({\bf k}',{\bf k})= V_{c}({\bf k^\prime},{\bf k})+i \sin{\theta}_{kk^\prime}
\; V_{s}({\bf k^\prime}, {\bf k}) \; {\bf \sigma}\cdot {\bf \hat{n}},
 \label{eq:a.3.3a}
\end{equation}
where ${\hat n}$ is the vector perpendicular to the scattering plane spanned by the
vectors {\bf k} and {\bf k'}, and $V_s = \frac{d}{dx} V_{ls}$ with $x=\cos
\theta_{kk^\prime}$. $V_{c}({\bf k^\prime},{\bf k})$ is the sum of the three
central interaction terms of Eq.~(\ref{eq:a.3.3a}).
For the partial wave decomposition one obtains (see e.g. Ref.~\cite{Landau:1990qp})
\begin{eqnarray}
 V_{c}({\bf k^\prime},{\bf k})&=& \frac{1}{4\pi} \sum_{l=0}^{\infty} 
\left[(l+1)V_{l_+}(k^\prime,k) +lV_{l_-}(k^\prime,k) \right] P_{l}(x)  \cr
V_{s}({\bf k^\prime},{\bf k})=  \frac{d}{dx} V_{ls}({\bf k^\prime},{\bf k}) &=& \frac{1}{4\pi}\frac{d}{dx} \sum_{l=0}^{\infty} 
\left[V_{l_+}(k^\prime,k)-V_{l_-} (k^\prime,k) \right] P_{l} (x).
\label{eq:a.3.3.b}
\end{eqnarray}
Using the standard orthogonality relations for the Legendre polynomials 
leads to the partial wave projected potential for a spin-0 + spin-1/2 system
\begin{eqnarray}
 V_{l_+}(k^\prime,k)&=&2\pi \int_{-1}^{1}dx P_{l}(x) V_{c}({\bf
 k^\prime},{\bf k})
 +2\pi l \int_{-1}^{1}dx P_{l}(x)V_{ls}({\bf k^\prime},{\bf k}) \cr
 V_{l_-}(k^\prime,k)&=&2\pi \int_{-1}^{1}dx P_{l}(x)V_{c}({\bf
 k^\prime},{\bf k})
 -2\pi (l+1) \int_{-1}^{1}dx P_{l}(x)V_{ls}({\bf k^\prime},{\bf k}) \; ,
\label{eq:a.3.3c}
\end{eqnarray}
where the indices refer to $j=l+\frac{1}{2}$ and $j=l-\frac{1}{2}$
respectively. 
The partial wave projected potential matrix enter a two-body LS equation, from which 
the standard scattering phase shifts are obtained. 

To test the quality of the momentum space representation of the CH89 optical
potential, we compare scattering phase shifts obtained in momentum space with those
independently calculated in coordinate
space using the techniques implemented in the coupled-channel code 
FRESCO~\cite{Thompson:1988zz}. This comparison is
given in Table~\ref{table-1} for n+$^{48}$Ca for projectile kinetic energies up to
50~MeV. The agreement is satisfactory, and gives confidence to use the momentum
space representation of the CH89 global optical potential as starting point for
constructing separable representations thereof.

%%%%%%%%%%%%%%%%%%%%%%%%%%%%%%%%%%%%%%%%%%%%%%%%%%%%%%%%%%%%%%%%%%%%%%%%%%%%%%%%%%%%

\begin{acknowledgments}
This work was performed in part under the
auspices of the U.~S.  Department of Energy under contract
No. DE-FG02-93ER40756 with Ohio University and DE-FG52-08NA28552 with Michigan State University;
under contract No. DE-SC0004084 and DE-SC0004087 (TORUS Collaboration), and by  Lawrence Livermore
National Laboratory under Contract DE-AC52-07NA27344. F.M. Nunes also acknowledges support from the National Science Foundation 
under grant PHY-0800026.  R.C. Johnson thanks the NSCL for their hospitality and support during his visits.
\end{acknowledgments}

%%%%%%%%%%%%%%%%%%%%%%%%%%%%%%%%%%%%%%%%%%%%%%%%%%%%%%%%%%%%%%%%%%%%

\bibliography{ESTdp}

\begin{thebibliography}{33}%
\makeatletter
\providecommand \@ifxundefined [1]{%
 \@ifx{#1\undefined}
}%
\providecommand \@ifnum [1]{%
 \ifnum #1\expandafter \@firstoftwo
 \else \expandafter \@secondoftwo
 \fi
}%
\providecommand \@ifx [1]{%
 \ifx #1\expandafter \@firstoftwo
 \else \expandafter \@secondoftwo
 \fi
}%
\providecommand \natexlab [1]{#1}%
\providecommand \enquote  [1]{``#1''}%
\providecommand \bibnamefont  [1]{#1}%
\providecommand \bibfnamefont [1]{#1}%
\providecommand \citenamefont [1]{#1}%
\providecommand \href@noop [0]{\@secondoftwo}%
\providecommand \href [0]{\begingroup \@sanitize@url \@href}%
\providecommand \@href[1]{\@@startlink{#1}\@@href}%
\providecommand \@@href[1]{\endgroup#1\@@endlink}%
\providecommand \@sanitize@url [0]{\catcode `\\12\catcode `\$12\catcode
  `\&12\catcode `\#12\catcode `\^12\catcode `\_12\catcode `\%12\relax}%
\providecommand \@@startlink[1]{}%
\providecommand \@@endlink[0]{}%
\providecommand \url  [0]{\begingroup\@sanitize@url \@url }%
\providecommand \@url [1]{\endgroup\@href {#1}{\urlprefix }}%
\providecommand \urlprefix  [0]{URL }%
\providecommand \Eprint [0]{\href }%
\providecommand \doibase [0]{http://dx.doi.org/}%
\providecommand \selectlanguage [0]{\@gobble}%
\providecommand \bibinfo  [0]{\@secondoftwo}%
\providecommand \bibfield  [0]{\@secondoftwo}%
\providecommand \translation [1]{[#1]}%
\providecommand \BibitemOpen [0]{}%
\providecommand \bibitemStop [0]{}%
\providecommand \bibitemNoStop [0]{.\EOS\space}%
\providecommand \EOS [0]{\spacefactor3000\relax}%
\providecommand \BibitemShut  [1]{\csname bibitem#1\endcsname}%
\let\auto@bib@innerbib\@empty
%</preamble>
\bibitem [{\citenamefont {Escher}\ \emph {et~al.}(2012)\citenamefont {Escher},
  \citenamefont {Burke}, \citenamefont {Dietrich}, \citenamefont {Scielzo},
  \citenamefont {Thompson},\ and\ \citenamefont {Younes}}]{RevModPhys.84.353}%
  \BibitemOpen
  \bibfield  {author} {\bibinfo {author} {\bibfnamefont {J.~E.}\ \bibnamefont
  {Escher}}, \bibinfo {author} {\bibfnamefont {J.~T.}\ \bibnamefont {Burke}},
  \bibinfo {author} {\bibfnamefont {F.~S.}\ \bibnamefont {Dietrich}}, \bibinfo
  {author} {\bibfnamefont {N.~D.}\ \bibnamefont {Scielzo}}, \bibinfo {author}
  {\bibfnamefont {I.~J.}\ \bibnamefont {Thompson}}, \ and\ \bibinfo {author}
  {\bibfnamefont {W.}~\bibnamefont {Younes}},\ }\href {\doibase
  10.1103/RevModPhys.84.353} {\bibfield  {journal} {\bibinfo  {journal} {Rev.
  Mod. Phys.}\ }\textbf {\bibinfo {volume} {84}},\ \bibinfo {pages} {353}
  (\bibinfo {year} {2012})}\BibitemShut {NoStop}%
\bibitem [{\citenamefont {Cizewski}\ \emph {et~al.}(2013)\citenamefont
  {Cizewski} \emph {et~al.}}]{jolie}%
  \BibitemOpen
  \bibfield  {author} {\bibinfo {author} {\bibfnamefont {J.}~\bibnamefont
  {Cizewski}} \emph {et~al.},\ }\href@noop {} {\bibfield  {journal} {\bibinfo
  {journal} {J. Phys. Conf}\ }\textbf {\bibinfo {volume} {420}},\ \bibinfo
  {pages} {012058} (\bibinfo {year} {2013})}\BibitemShut {NoStop}%
\bibitem [{\citenamefont {Kozub}\ \emph {et~al.}(2012)\citenamefont {Kozub},
  \citenamefont {Arbanas}, \citenamefont {Adekola}, \citenamefont {Bardayan},
  \citenamefont {Blackmon} \emph {et~al.}}]{Kozub:2012ka}%
  \BibitemOpen
  \bibfield  {author} {\bibinfo {author} {\bibfnamefont {R.}~\bibnamefont
  {Kozub}}, \bibinfo {author} {\bibfnamefont {G.}~\bibnamefont {Arbanas}},
  \bibinfo {author} {\bibfnamefont {A.}~\bibnamefont {Adekola}}, \bibinfo
  {author} {\bibfnamefont {D.}~\bibnamefont {Bardayan}}, \bibinfo {author}
  {\bibfnamefont {J.}~\bibnamefont {Blackmon}},  \emph {et~al.},\ }\href
  {\doibase 10.1103/PhysRevLett.109.172501} {\bibfield  {journal} {\bibinfo
  {journal} {Phys. Rev. Lett.}\ }\textbf {\bibinfo {volume} {109}},\ \bibinfo
  {pages} {172501} (\bibinfo {year} {2012})}\BibitemShut {NoStop}%
%%CITATION = PRLTA,109,172501;%%
\bibitem [{\citenamefont {Schmitt}\ \emph {et~al.}(2012)\citenamefont
  {Schmitt}, \citenamefont {Jones}, \citenamefont {Bey}, \citenamefont {Ahn},
  \citenamefont {Bardayan} \emph {et~al.}}]{Schmitt:2012bt}%
  \BibitemOpen
  \bibfield  {author} {\bibinfo {author} {\bibfnamefont {K.}~\bibnamefont
  {Schmitt}}, \bibinfo {author} {\bibfnamefont {K.}~\bibnamefont {Jones}},
  \bibinfo {author} {\bibfnamefont {A.}~\bibnamefont {Bey}}, \bibinfo {author}
  {\bibfnamefont {S.}~\bibnamefont {Ahn}}, \bibinfo {author} {\bibfnamefont
  {D.}~\bibnamefont {Bardayan}},  \emph {et~al.},\ }\href {\doibase
  10.1103/PhysRevLett.108.192701} {\bibfield  {journal} {\bibinfo  {journal}
  {Phys. Rev. Lett.}\ }\textbf {\bibinfo {volume} {108}},\ \bibinfo {pages}
  {192701} (\bibinfo {year} {2012})}\BibitemShut {NoStop}%
%%CITATION = ARXIV:1203.3081;%%
\bibitem [{\citenamefont {Jones}\ \emph {et~al.}(2011)\citenamefont {Jones},
  \citenamefont {Nunes}, \citenamefont {Adekola}, \citenamefont {Bardayan},
  \citenamefont {Blackmon} \emph {et~al.}}]{Jones:2011kp}%
  \BibitemOpen
  \bibfield  {author} {\bibinfo {author} {\bibfnamefont {K.}~\bibnamefont
  {Jones}}, \bibinfo {author} {\bibfnamefont {F.}~\bibnamefont {Nunes}},
  \bibinfo {author} {\bibfnamefont {A.}~\bibnamefont {Adekola}}, \bibinfo
  {author} {\bibfnamefont {D.}~\bibnamefont {Bardayan}}, \bibinfo {author}
  {\bibfnamefont {J.}~\bibnamefont {Blackmon}},  \emph {et~al.},\ }\href
  {\doibase 10.1103/PhysRevC.84.034601} {\bibfield  {journal} {\bibinfo
  {journal} {Phys. Rev.}\ }\textbf {\bibinfo {volume} {C84}},\ \bibinfo {pages}
  {034601} (\bibinfo {year} {2011})}\BibitemShut {NoStop}%
%%CITATION = ARXIV:1105.4755;%%
\bibitem [{\citenamefont {Varner}\ \emph {et~al.}(1991)\citenamefont {Varner},
  \citenamefont {Thompson}, \citenamefont {McAbee}, \citenamefont {Ludwig},\
  and\ \citenamefont {Clegg}}]{Varner:1991zz}%
  \BibitemOpen
  \bibfield  {author} {\bibinfo {author} {\bibfnamefont {R.}~\bibnamefont
  {Varner}}, \bibinfo {author} {\bibfnamefont {W.}~\bibnamefont {Thompson}},
  \bibinfo {author} {\bibfnamefont {T.}~\bibnamefont {McAbee}}, \bibinfo
  {author} {\bibfnamefont {E.}~\bibnamefont {Ludwig}}, \ and\ \bibinfo {author}
  {\bibfnamefont {T.}~\bibnamefont {Clegg}},\ }\href {\doibase 10.10
  16/0370-1573(91)90039-O} {\bibfield  {journal} {\bibinfo  {journal}
  {Phys. Rept.}\ }\textbf {\bibinfo {volume} {201}},\ \bibinfo {pages} {57}
  (\bibinfo {year} {1991})}\BibitemShut {NoStop}%
%%CITATION = PRPLC,201,57;%%
\bibitem [{\citenamefont {Weppner}\ \emph {et~al.}(2009)\citenamefont
  {Weppner}, \citenamefont {Penney}, \citenamefont {Diffendale},\ and\
  \citenamefont {Vittorini}}]{Weppner:2009qy}%
  \BibitemOpen
  \bibfield  {author} {\bibinfo {author} {\bibfnamefont {S.}~\bibnamefont
  {Weppner}}, \bibinfo {author} {\bibfnamefont {R.}~\bibnamefont {Penney}},
  \bibinfo {author} {\bibfnamefont {G.}~\bibnamefont {Diffendale}}, \ and\
  \bibinfo {author} {\bibfnamefont {G.}~\bibnamefont {Vittorini}},\ }\href
  {\doibase 10.1103/PhysRevC.80.034608} {\bibfield  {journal} {\bibinfo
  {journal} {Phys. Rev.}\ }\textbf {\bibinfo {volume} {C80}},\ \bibinfo {pages}
  {034608} (\bibinfo {year} {2009})}\BibitemShut {NoStop}%
%%CITATION = ARXIV:0906.3758;%%
\bibitem [{\citenamefont {Koning}\ and\ \citenamefont
  {Delaroche}(2003)}]{Koning:2003zz}%
  \BibitemOpen
  \bibfield  {author} {\bibinfo {author} {\bibfnamefont {A.}~\bibnamefont
  {Koning}}\ and\ \bibinfo {author} {\bibfnamefont {J.}~\bibnamefont
  {Delaroche}},\ }\href@noop {} {\bibfield  {journal} {\bibinfo  {journal}
  {Nucl. Phys.}\ }\textbf {\bibinfo {volume} {A713}},\ \bibinfo {pages} {231}
  (\bibinfo {year} {2003})}\BibitemShut {NoStop}%
%%CITATION = NUPHA,A713,231;%%
\bibitem [{\citenamefont {Becchetti}\ and\ \citenamefont
  {Greenlees}(1969)}]{Becchetti:1969zz}%
  \BibitemOpen
  \bibfield  {author} {\bibinfo {author} {\bibfnamefont {J.}~\bibnamefont
  {Becchetti}, \bibfnamefont {F.D.}}\ and\ \bibinfo {author} {\bibfnamefont
  {G.}~\bibnamefont {Greenlees}},\ }\href {\doibase 10.1103/PhysRev.182.1190}
  {\bibfield  {journal} {\bibinfo  {journal} {Phys. Rev.}\ }\textbf {\bibinfo
  {volume} {182}},\ \bibinfo {pages} {1190} (\bibinfo {year}
  {1969})}\BibitemShut {NoStop}%
%%CITATION = PHRVA,182,1190;%%
\bibitem [{\citenamefont {Deltuva}\ and\ \citenamefont
  {Fonseca}(2009)}]{Deltuva:2009fp}%
  \BibitemOpen
  \bibfield  {author} {\bibinfo {author} {\bibfnamefont {A.}~\bibnamefont
  {Deltuva}}\ and\ \bibinfo {author} {\bibfnamefont {A.}~\bibnamefont
  {Fonseca}},\ }\href {\doibase 10.1103/PhysRevC.79.014606} {\bibfield
  {journal} {\bibinfo  {journal} {Phys. Rev.}\ }\textbf {\bibinfo {volume}
  {C79}},\ \bibinfo {pages} {014606} (\bibinfo {year} {2009})}\BibitemShut
  {NoStop}%
%%CITATION = ARXIV:0901.0875;%%
\bibitem [{\citenamefont {Deltuva}(2009)}]{Deltuva:2009cr}%
  \BibitemOpen
  \bibfield  {author} {\bibinfo {author} {\bibfnamefont {A.}~\bibnamefont
  {Deltuva}},\ }\href {\doibase 10.1103/PhysRevC.79.054603} {\bibfield
  {journal} {\bibinfo  {journal} {Phys. Rev.}\ }\textbf {\bibinfo {volume}
  {C79}},\ \bibinfo {pages} {054603} (\bibinfo {year} {2009})}\BibitemShut
  {NoStop}%
%%CITATION = ARXIV:0903.3721;%%
\bibitem [{\citenamefont {Nunes}\ and\ \citenamefont
  {Deltuva}(2011)}]{Nunes:2011cv}%
  \BibitemOpen
  \bibfield  {author} {\bibinfo {author} {\bibfnamefont {F.}~\bibnamefont
  {Nunes}}\ and\ \bibinfo {author} {\bibfnamefont {A.}~\bibnamefont
  {Deltuva}},\ }\href {\doibase 10.1103/PhysRevC.84.034607} {\bibfield
  {journal} {\bibinfo  {journal} {Phys. Rev.}\ }\textbf {\bibinfo {volume}
  {C84}},\ \bibinfo {pages} {034607} (\bibinfo {year} {2011})}\BibitemShut
  {NoStop}%
%%CITATION = ARXIV:1108.2519;%%
\bibitem [{\citenamefont {Upadhyay}\ \emph {et~al.}(2012)\citenamefont
  {Upadhyay}, \citenamefont {Deltuva},\ and\ \citenamefont
  {Nunes}}]{Upadhyay:2011ta}%
  \BibitemOpen
  \bibfield  {author} {\bibinfo {author} {\bibfnamefont {N.}~\bibnamefont
  {Upadhyay}}, \bibinfo {author} {\bibfnamefont {A.}~\bibnamefont {Deltuva}}, \
  and\ \bibinfo {author} {\bibfnamefont {F.}~\bibnamefont {Nunes}},\ }\href
  {\doibase 10.1103/PhysRevC.85.054621} {\bibfield  {journal} {\bibinfo
  {journal} {Phys. Rev.}\ }\textbf {\bibinfo {volume} {C85}},\ \bibinfo {pages}
  {054621} (\bibinfo {year} {2012})},\ \Eprint {http://arxiv.org/abs/1112.5338}
  {arXiv:1112.5338 [nucl-th]} \BibitemShut {NoStop}%
%%CITATION = ARXIV:1112.5338;%%
\bibitem [{\citenamefont {Mukhamedzhanov}\ \emph {et~al.}(2012)\citenamefont
  {Mukhamedzhanov}, \citenamefont {Eremenko},\ and\ \citenamefont
  {Sattarov}}]{Mukhamedzhanov:2012qv}%
  \BibitemOpen
  \bibfield  {author} {\bibinfo {author} {\bibfnamefont {A.}~\bibnamefont
  {Mukhamedzhanov}}, \bibinfo {author} {\bibfnamefont {V.}~\bibnamefont
  {Eremenko}}, \ and\ \bibinfo {author} {\bibfnamefont {A.}~\bibnamefont
  {Sattarov}},\ }\href {\doibase 10.1103/PhysRevC.86.034001} {\bibfield
  {journal} {\bibinfo  {journal} {Phys. Rev.}\ }\textbf {\bibinfo {volume}
  {C86}},\ \bibinfo {pages} {034001} (\bibinfo {year} {2012})}\BibitemShut
  {NoStop}%
%%CITATION = ARXIV:1206.3791;%%
\bibitem [{\citenamefont {Haidenbauer}\ and\ \citenamefont
  {Plessas}(1983)}]{Haidenbauer:1982if}%
  \BibitemOpen
  \bibfield  {author} {\bibinfo {author} {\bibfnamefont {J.}~\bibnamefont
  {Haidenbauer}}\ and\ \bibinfo {author} {\bibfnamefont {W.}~\bibnamefont
  {Plessas}},\ }\href {\doibase 10.1103/PhysRevC.27.63} {\bibfield  {journal}
  {\bibinfo  {journal} {Phys. Rev.}\ }\textbf {\bibinfo {volume} {C27}},\
  \bibinfo {pages} {63} (\bibinfo {year} {1983})}\BibitemShut {NoStop}%
%%CITATION = PHRVA,C27,63;%%
\bibitem [{\citenamefont {Haidenbauer}\ \emph {et~al.}(1986)\citenamefont
  {Haidenbauer}, \citenamefont {Koike},\ and\ \citenamefont
  {Plessas}}]{Haidenbauer:1986zza}%
  \BibitemOpen
  \bibfield  {author} {\bibinfo {author} {\bibfnamefont {J.}~\bibnamefont
  {Haidenbauer}}, \bibinfo {author} {\bibfnamefont {Y.}~\bibnamefont {Koike}},
  \ and\ \bibinfo {author} {\bibfnamefont {W.}~\bibnamefont {Plessas}},\ }\href
  {\doibase 10.1103/PhysRevC.33.439} {\bibfield  {journal} {\bibinfo  {journal}
  {Phys. Rev.}\ }\textbf {\bibinfo {volume} {C33}},\ \bibinfo {pages} {439}
  (\bibinfo {year} {1986})}\BibitemShut {NoStop}%
%%CITATION = PHRVA,C33,439;%%
\bibitem [{\citenamefont {Berthold}\ \emph {et~al.}(1990)\citenamefont
  {Berthold}, \citenamefont {Stadler},\ and\ \citenamefont
  {Zankel}}]{Berthold:1990zz}%
  \BibitemOpen
  \bibfield  {author} {\bibinfo {author} {\bibfnamefont {G.}~\bibnamefont
  {Berthold}}, \bibinfo {author} {\bibfnamefont {A.}~\bibnamefont {Stadler}}, \
  and\ \bibinfo {author} {\bibfnamefont {H.}~\bibnamefont {Zankel}},\ }\href
  {\doibase 10.1103/PhysRevC.41.1365} {\bibfield  {journal} {\bibinfo
  {journal} {Phys. Rev.}\ }\textbf {\bibinfo {volume} {C41}},\ \bibinfo {pages}
  {1365} (\bibinfo {year} {1990})}\BibitemShut {NoStop}%
%%CITATION = PHRVA,C41,1365;%%
\bibitem [{\citenamefont {Schnizer}\ and\ \citenamefont
  {Plessas}(1990)}]{Schnizer:1990gf}%
  \BibitemOpen
  \bibfield  {author} {\bibinfo {author} {\bibfnamefont {W.}~\bibnamefont
  {Schnizer}}\ and\ \bibinfo {author} {\bibfnamefont {W.}~\bibnamefont
  {Plessas}},\ }\href {\doibase 10.1103/PhysRevC.41.1095} {\bibfield  {journal}
  {\bibinfo  {journal} {Phys. Rev.}\ }\textbf {\bibinfo {volume} {C41}},\
  \bibinfo {pages} {1095} (\bibinfo {year} {1990})}\BibitemShut {NoStop}%
%%CITATION = PHRVA,C41,1095;%%
\bibitem [{\citenamefont {Entem}\ \emph {et~al.}(2001)\citenamefont {Entem},
  \citenamefont {Fernandez},\ and\ \citenamefont {Valcarce}}]{Entem:2001it}%
  \BibitemOpen
  \bibfield  {author} {\bibinfo {author} {\bibfnamefont {D.}~\bibnamefont
  {Entem}}, \bibinfo {author} {\bibfnamefont {F.}~\bibnamefont {Fernandez}}, \
  and\ \bibinfo {author} {\bibfnamefont {A.}~\bibnamefont {Valcarce}},\ }\href
  {\doibase 10.1088/0954-3899/27/7/312} {\bibfield  {journal} {\bibinfo
  {journal} {J. Phys.}\ }\textbf {\bibinfo {volume} {G27}},\ \bibinfo {pages}
  {1537} (\bibinfo {year} {2001})}\BibitemShut {NoStop}%
%%CITATION = INSPIRE-561051;%%
\bibitem [{\citenamefont {Ueda}\ and\ \citenamefont
  {Ikegami}(1994)}]{Ueda:1994ur}%
  \BibitemOpen
  \bibfield  {author} {\bibinfo {author} {\bibfnamefont {T.}~\bibnamefont
  {Ueda}}\ and\ \bibinfo {author} {\bibfnamefont {Y.}~\bibnamefont {Ikegami}},\
  }\href {\doibase 10.1143/PTP.91.85} {\bibfield  {journal} {\bibinfo
  {journal} {Prog. Theor. Phys.}\ }\textbf {\bibinfo {volume} {91}},\ \bibinfo
  {pages} {85} (\bibinfo {year} {1994})}\BibitemShut {NoStop}%
%%CITATION = PTPKA,91,85;%%
\bibitem [{\citenamefont {Gal}\ and\ \citenamefont
  {Garcilazo}(2011)}]{Gal:2011yp}%
  \BibitemOpen
  \bibfield  {author} {\bibinfo {author} {\bibfnamefont {A.}~\bibnamefont
  {Gal}}\ and\ \bibinfo {author} {\bibfnamefont {H.}~\bibnamefont
  {Garcilazo}},\ }\href {\doibase 10.1016/j.nuclphysa.2011.06.022} {\bibfield
  {journal} {\bibinfo  {journal} {Nucl. Phys.}\ }\textbf {\bibinfo {volume}
  {A864}},\ \bibinfo {pages} {153} (\bibinfo {year} {2011})}\BibitemShut
  {NoStop}%
%%CITATION = ARXIV:1103.4757;%%
\bibitem [{\citenamefont {Ghovanlou}\ and\ \citenamefont
  {Lehman}(1974)}]{Ghovanlou:1974zza}%
  \BibitemOpen
  \bibfield  {author} {\bibinfo {author} {\bibfnamefont {A.}~\bibnamefont
  {Ghovanlou}}\ and\ \bibinfo {author} {\bibfnamefont {D.~R.}\ \bibnamefont
  {Lehman}},\ }\href {\doibase 10.1103/PhysRevC.9.1730} {\bibfield  {journal}
  {\bibinfo  {journal} {Phys. Rev.}\ }\textbf {\bibinfo {volume} {C9}},\
  \bibinfo {pages} {1730} (\bibinfo {year} {1974})}\BibitemShut {NoStop}%
%%CITATION = PHRVA,C9,1730;%%
\bibitem [{\citenamefont {Eskandarian}\ and\ \citenamefont
  {Afnan}(1992)}]{Eskandarian:1992zz}%
  \BibitemOpen
  \bibfield  {author} {\bibinfo {author} {\bibfnamefont {A.}~\bibnamefont
  {Eskandarian}}\ and\ \bibinfo {author} {\bibfnamefont {I.~R.}\ \bibnamefont
  {Afnan}},\ }\href {\doibase 10.1103/PhysRevC.46.2344} {\bibfield  {journal}
  {\bibinfo  {journal} {Phys. Rev.}\ }\textbf {\bibinfo {volume} {C46}},\
  \bibinfo {pages} {2344} (\bibinfo {year} {1992})}\BibitemShut {NoStop}%
%%CITATION = PHRVA,C46,2344;%%
\bibitem [{\citenamefont {Miyagawa}\ and\ \citenamefont
  {Koike}(1989)}]{MiyagawaK}%
  \BibitemOpen
  \bibfield  {author} {\bibinfo {author} {\bibfnamefont {K.}~\bibnamefont
  {Miyagawa}}\ and\ \bibinfo {author} {\bibfnamefont {Y.}~\bibnamefont
  {Koike}},\ }\href@noop {} {\bibfield  {journal} {\bibinfo  {journal} {Prog.
  Theor. Phys.}\ }\textbf {\bibinfo {volume} {82}},\ \bibinfo {pages} {329}
  (\bibinfo {year} {1989})}\BibitemShut {NoStop}%
%%CITATION = PHRVA,C46,2344;%%
\bibitem [{\citenamefont {Ernst}\ \emph {et~al.}(1973)\citenamefont {Ernst},
  \citenamefont {Shakin},\ and\ \citenamefont {Thaler}}]{Ernst:1973zzb}%
  \BibitemOpen
  \bibfield  {author} {\bibinfo {author} {\bibfnamefont {D.}~\bibnamefont
  {Ernst}}, \bibinfo {author} {\bibfnamefont {C.}~\bibnamefont {Shakin}}, \
  and\ \bibinfo {author} {\bibfnamefont {R.}~\bibnamefont {Thaler}},\ }\href
  {\doibase 10.1103/PhysRevC.8.46} {\bibfield  {journal} {\bibinfo  {journal}
  {Phys. Rev.}\ }\textbf {\bibinfo {volume} {C8}},\ \bibinfo {pages} {46}
  (\bibinfo {year} {1973})}\BibitemShut {NoStop}%
%%CITATION = PHRVA,C8,46;%%
\bibitem [{\citenamefont {Ernst}\ \emph {et~al.}(1974)\citenamefont {Ernst},
  \citenamefont {Londergan}, \citenamefont {Moniz},\ and\ \citenamefont
  {Thaler}}]{Ernst:1974zzb}%
  \BibitemOpen
  \bibfield  {author} {\bibinfo {author} {\bibfnamefont {D.}~\bibnamefont
  {Ernst}}, \bibinfo {author} {\bibfnamefont {J.}~\bibnamefont {Londergan}},
  \bibinfo {author} {\bibfnamefont {E.}~\bibnamefont {Moniz}}, \ and\ \bibinfo
  {author} {\bibfnamefont {R.}~\bibnamefont {Thaler}},\ }\href {\doibase
  10.1103/PhysRevC.10.1708} {\bibfield  {journal} {\bibinfo  {journal}
  {Phys. Rev.}\ }\textbf {\bibinfo {volume} {C10}},\ \bibinfo {pages} {1708}
  (\bibinfo {year} {1974})}\BibitemShut {NoStop}%
%%CITATION = PHRVA,C10,1708;%%
\bibitem [{\citenamefont {Pearce}(1987)}]{Pearce:1987zz}%
  \BibitemOpen
  \bibfield  {author} {\bibinfo {author} {\bibfnamefont {B.}~\bibnamefont
  {Pearce}},\ }\href {\doibase 10.1103/PhysRevC.36.471} {\bibfield  {journal}
  {\bibinfo  {journal} {Phys. Rev.}\ }\textbf {\bibinfo {volume} {C36}},\
  \bibinfo {pages} {471} (\bibinfo {year} {1987})}\BibitemShut {NoStop}%
%%CITATION = PHRVA,C36,471;%%
\bibitem [{\citenamefont {Huber}\ \emph {et~al.}(1997)\citenamefont {Huber},
  \citenamefont {Witala}, \citenamefont {Nogga}, \citenamefont {Gloeckle},\
  and\ \citenamefont {Kamada}}]{Huber:1996td}%
  \BibitemOpen
  \bibfield  {author} {\bibinfo {author} {\bibfnamefont {D.}~\bibnamefont
  {Huber}}, \bibinfo {author} {\bibfnamefont {H.}~\bibnamefont {Witala}},
  \bibinfo {author} {\bibfnamefont {A.}~\bibnamefont {Nogga}}, \bibinfo
  {author} {\bibfnamefont {W.}~\bibnamefont {Gloeckle}}, \ and\ \bibinfo
  {author} {\bibfnamefont {H.}~\bibnamefont {Kamada}},\ }\href {\doibase
  10.1007/s006010050057} {\bibfield  {journal} {\bibinfo  {journal} {Few Body
  Syst.}\ }\textbf {\bibinfo {volume} {22}},\ \bibinfo {pages} {107} (\bibinfo
  {year} {1997})}\BibitemShut {NoStop}%
%%CITATION = NUCL-TH/9611021;%%
\bibitem [{\citenamefont {Thompson}(1988)}]{Thompson:1988zz}%
  \BibitemOpen
  \bibfield  {author} {\bibinfo {author} {\bibfnamefont {I.~J.}\ \bibnamefont
  {Thompson}},\ }\href@noop {} {\bibfield  {journal} {\bibinfo  {journal}
  {Comput. Phys. Rept.}\ }\textbf {\bibinfo {volume} {7}},\ \bibinfo {pages}
  {167} (\bibinfo {year} {1988})}\BibitemShut {NoStop}%
%%CITATION = CPHRE,7,167;%%
\bibitem [{\citenamefont {Lovelace}(1964)}]{Lovelace:1964mq}%
  \BibitemOpen
  \bibfield  {author} {\bibinfo {author} {\bibfnamefont {C.}~\bibnamefont
  {Lovelace}},\ }\href {\doibase 10.1103/PhysRev.135.B1225} {\bibfield
  {journal} {\bibinfo  {journal} {Phys. Rev.}\ }\textbf {\bibinfo {volume}
  {135}},\ \bibinfo {pages} {B1225} (\bibinfo {year} {1964})}\BibitemShut
  {NoStop}%
%%CITATION = PHRVA,135,B1225;%%
\bibitem [{\citenamefont {Bogner}\ \emph {et~al.}(2007)\citenamefont {Bogner},
  \citenamefont {Furnstahl},\ and\ \citenamefont {Perry}}]{Bogner:2006pc}%
  \BibitemOpen
  \bibfield  {author} {\bibinfo {author} {\bibfnamefont {S.}~\bibnamefont
  {Bogner}}, \bibinfo {author} {\bibfnamefont {R.}~\bibnamefont {Furnstahl}}, \
  and\ \bibinfo {author} {\bibfnamefont {R.}~\bibnamefont {Perry}},\ }\href
  {\doibase 10.1103/PhysRevC.75.061001} {\bibfield  {journal} {\bibinfo
  {journal} {Phys. Rev.}\ }\textbf {\bibinfo {volume} {C75}},\ \bibinfo {pages}
  {061001} (\bibinfo {year} {2007})}\BibitemShut {NoStop}%
%%CITATION = NUCL-TH/0611045;%%
\bibitem [{\citenamefont {Rodriguez-Gallardo}\ \emph
  {et~al.}(2008)\citenamefont {Rodriguez-Gallardo}, \citenamefont {Deltuva},
  \citenamefont {Cravo}, \citenamefont {Crespo},\ and\ \citenamefont
  {Fonseca}}]{RodriguezGallardo:2007dc}%
  \BibitemOpen
  \bibfield  {author} {\bibinfo {author} {\bibfnamefont {M.}~\bibnamefont
  {Rodriguez-Gallardo}}, \bibinfo {author} {\bibfnamefont {A.}~\bibnamefont
  {Deltuva}}, \bibinfo {author} {\bibfnamefont {E.}~\bibnamefont {Cravo}},
  \bibinfo {author} {\bibfnamefont {R.}~\bibnamefont {Crespo}}, \ and\ \bibinfo
  {author} {\bibfnamefont {A.}~\bibnamefont {Fonseca}},\ }\href {\doibase
  10.1103/PhysRevC.78.034602} {\bibfield  {journal} {\bibinfo  {journal}
  {Phys. Rev.}\ }\textbf {\bibinfo {volume} {C78}},\ \bibinfo {pages} {034602}
  (\bibinfo {year} {2008})}\BibitemShut {NoStop}%
%%CITATION = ARXIV:0712.1292;%%
\bibitem [{\citenamefont {Landau}(2007)}]{Landau:1990qp}%
  \BibitemOpen
  \bibfield  {author} {\bibinfo {author} {\bibfnamefont {R.}~\bibnamefont
  {Landau}},\ }\href@noop {} {\bibfield  {journal} {\bibinfo  {journal}
  {{Quantum mechanics II: A second course in quantum theory}}\ }\textbf
  {\bibinfo {volume} {Wiley-VCH}} (\bibinfo {year} {2007})}\BibitemShut
  {NoStop}%
%%CITATION = INSPIRE-306191;%%
\end{thebibliography}%

\clearpage
%%%%%%%%%%%%%%%%%%%%%%%%%%%%%%%%%%%%%%%%%%%%%%%%%%%%%%

\begin{table}
\begin{tabular}{|c|c|c|c| }
\hline\hline
   system    & partial wave(s)  & rank & EST support point(s) [MeV]\\
\hline
             &  $l\ge 10$       &  1   & 40         \\
n$+^{48}$Ca  &  $l\ge  8$       &  2   & 29, 47      \\
             &  $l\ge  6$       &  3   & 16, 36, 47      \\
             &  $l\ge  0$       &  4   & 6,  15, 36, 47   \\[1ex]
\hline
%             &                   &      &                \\
             &  $l\ge  16$       &  1   & 40 \\
n$+^{132}$Sn &  $l\ge  13$       &  2   & 35, 48   \\
and 	     &  $l\ge  11$       &  3   & 24, 39, 48       \\
n$+^{208}$Pb &  $l\ge   6$       &  4   & 11, 21, 36, 45          \\
             &  $\mathbf {l\ge   0}$       &  $\mathbf{5}$   & $\mathbf{ 5, 11, 21, 36, 47}$
\\[1ex]
\hline
\end{tabular}
\caption{The EST support points at c.m. energies $E_{k_i}$
used  for constructing the separable
representation of the partial wave s-matrix of the n$+^{48}$Ca
 and n$+^{208}$Pb systems.
The support points in the last row for the n$+^{208}$Pb system
given in bold face indicate the universal set of support points, which can be used to
construct a representation for all nuclei given by the
CH89~\cite{Varner:1991zz} phenomenological optical potential.
}
\label{table-4}
\end{table}

\vspace{10mm}

\begin{table}
\begin{tabular}{|c|cccc|}
\hline\hline
   $n_{max}$ &     $S_1^R$    &   $S_1^I$   & $S_2^R$   & $S_2^I$    \\
   \hline
   0 & -4.000000  &  0.000000   &  0.000000 &   2.000000   \\
   1 & -4.023898  &  0.031864   &  0.039830 &   1.980085    \\
   2 & -4.023383  &  0.031589   &  0.038663 &   1.980377    \\
   3 & -4.023396  &  0.031594   &  0.038703 &   1.980370   \\
\hline \hline
\end{tabular}
 \caption{Numerical values of the sums of Eqs~(\ref{eq:a.2.6}) evaluated for $\rho=0.5$ and $
\alpha= 5.8$ (corresponding to $A=$ 40): results for CH89~\cite{Varner:1991zz}. The series is
summed up to the value of $n_{max}$ listed in the first column.
}
\label{table-2}
\end{table}

\vspace{10mm}

\begin{table}
\begin{tabular}{|c|cccc|}
\hline\hline
   $n_{max}$ &     $S_1^R$    &   $S_1^I$   & $S_2^R$   & $S_2^I$    \\
   \hline
   0 & -4.000000  &  0.000000   &  0.000000 &   2.000000   \\
   1 & -4.001453 &   0.031864   &  0.039830  &   1.998789  \\
   2 & -4.001451  &  0.031589   &  0.038663  &  1.998790   \\
   3 & -4.001451 &   0.031594   &  0.038703  &  1.998790   \\
   4 & -4.023395  &  0.031593   &  0.038702  & 1.980370    \\
   5 & -4.023395  &  0.031593   &  0.038702  &  1.980370    \\
   \hline\hline
  \end{tabular}
   \caption{Numerical values of the sums of Eqs~(\ref{eq:a.2.6}) evaluated for $\rho= 0.5$
and $ \alpha= 3.0$ (corresponding to $A=$ 12): results for WP~\cite{Weppner:2009qy}. The series is
summed up to the value of $n_{max}$ listed in the first column.
}
\label{table-3}
\end{table}

\vspace{10mm}

\begin{table}
\begin{tabular}{|c|cccc| }
\hline\hline
   $E_{lab}$ [MeV] &&&$ \rm \delta_{l=1,j=\frac{3}{2}} [deg]  $&\\
\hline
 & & k-space &&  r-space \\
\hline
  5 & &( -73.78 ,13.75 )& & (-73.84 ,13.75)   \\
  10 && (66.24 , 17.02)& &(66.18 ,17.00) \\
  20 && (18.09 , 19.51) & &(18.02 ,19.45) \\
  40 && (-38.03 , 22.48) & &(-38.08 , 22.47) \\
  50 && (-57.83 , 23.48) & &(-57.88 ,23.47) \\[1ex]
\hline \hline
\end{tabular}
\caption{The p-wave phase shift $\delta_{l=1,j=\frac{3}{2}}$ calculated from the
CH89~\cite{Varner:1991zz} optical potential for n+$^{48}$Ca elastic scattering as function of the
projectile laboratory energy. The second column shows phase shifts computed
in momentum space, while the third column gives the coordinate space calculation
based on the coupled-channel code FRESCO~\cite{Thompson:1988zz}.
}
\label{table-1}
\end{table}

\clearpage
%%%%%%%%%%%%%%%%%%%%%%%%%%%%%%%%%%%%%%%%%%%%%%%%%%%%%%

\newpage

\noindent

\noindent

\noindent

\noindent

\noindent

\noindent

\noindent

\end{document}